\documentclass[prd,superscriptaddress, nofootinbib, amsmath, amssymb,aps, floatfix, 11pt]{revtex4-1}

\usepackage{graphicx,xcolor,setspace}
\usepackage[
    colorlinks=true,
    linkcolor=blue,
    urlcolor=blue,
    citecolor=blue
]{hyperref}
\usepackage{subcaption}
\usepackage[capitalise]{cleveref}
\usepackage{siunitx}
\usepackage{color,epsfig}
\usepackage{ifpdf}
\usepackage{amsmath}
\usepackage{bm}
\usepackage[english]{babel}
\usepackage{amsfonts}%
\usepackage{amssymb}
\usepackage{braket}
\usepackage{hyperref}
\usepackage{enumerate}
\usepackage{cancel}
\usepackage{multirow}
\usepackage{tikz}
\usepackage[compat=1.0.0]{tikz-feynman}
\usepackage{booktabs}
\usepackage{ulem}
\usepackage{array}
\usepackage{ulem,fancyvrb}
\usepackage{xcolor}
\usepackage{dcolumn}
\usepackage{bm}
\usepackage{slashed}

\newcommand{\GeV}{\rm{GeV}}

\definecolor{nicered}{rgb}{0.7,0.1,0.1}
\definecolor{nicegreen}{rgb}{0.1,0.5,0.1}
\definecolor{maroon}{cmyk}{0,0.87,0.68,0.32}
\hypersetup{colorlinks,citecolor= nicegreen,linkcolor= nicered}

\def\comments{true}

\ifx\comments\undefined
	\newcommand{\comment}[1]{}
\else
	\newcommand{\comment}[1]{#1}
\fi

\begin{document}

\title{Probing the Scotogenic Dirac Model with FIMP Dark Matter and $\Delta N_{\rm eff}$}

\author{Shu-Yuan Guo}
\email{shyuanguo@ytu.edu.cn}
\affiliation{Department of Physics, Yantai University, Yantai 264005, China}

\author{Man-Yu Zhao}
\email{manyuzhao@s.ytu.edu.cn}
\affiliation{Department of Physics, Yantai University, Yantai 264005, China}

\begin{abstract}
We study a feebly interacting massive particle realization of the Scotogenic Dirac Model in which the lightest neutral fermion $N_1$ serves as a dark matter candidate, produced via the freeze-in or super-WIMP mechanism. The model generates Dirac neutrino masses at one loop, resulting in a rank-2 mass matrix that predicts one nearly massless neutrino. We analyze the DM relic density for various next-to-lightest odd particles (NLOPs), finding that coannihilation effects and enhanced annihilation channels are crucial for achieving the correct thermal freeze-out abundance of the NLOP.
We provide a detailed analysis of the model's implications for the effective number of relativistic species, $\Delta N_{\mathrm{eff}}$, which receives contributions from both a thermal bath of right-handed neutrinos and non-thermal energy injection due to late NLOP decays. Through an extensive parameter scan, we identify viable parameter space for all NLOP candidates that satisfies constraints from DM relic density, lepton flavor violation, Big Bang Nucleosynthesis, Cosmic Microwave Background, and $\Delta N_{\mathrm{eff}}$.
\end{abstract}

\maketitle

\section{Introduction}
The phenomenon of neutrino oscillations~\cite{Davis:1968cp,Super-Kamiokande:1998kpq,DayaBay:2012fng,RENO:2012mkc,K2K:2004iot,DoubleChooz:2011ymz} provides strong evidence that neutrinos must have non-zero masses. For neutral fermions such as neutrinos, two types of mass are theoretically possible: Dirac and Majorana. On the experimental front, extensive efforts have been devoted to searching for lepton number violation—a featured signature of Majorana neutrinos. Experiments at colliders~\cite{Cai:2017mow,CMS:2018jxx} and neutrinoless-double-beta-decay ($0\nu\beta\beta$)~\cite{Dolinski:2019nrj,EXO-200:2019rkq,LEGEND:2021bnm,Majorana:2019nbd,CUPID:2020aow,KamLAND-Zen:2022tow} searches, in particular, have so far yielded no positive signals. As a result, the alternative scenario has received increasing attention in recent years~\cite{Yao:2018ekp,CentellesChulia:2018bkz,Calle:2018ovc,Jana:2019mgj,Saad:2019bqf,Farzan:2012sa,Gu:2006dc,Gu:2007ug,Chulia:2016ngi,Bonilla:2016diq,Wang:2016lve,Borah:2017leo,Wang:2017mcy,CentellesChulia:2017koy,Ma:2017kgb,Yao:2017vtm,Bonilla:2017ekt,Ibarra:2017tju,Borah:2017dmk,Das:2017ski,Yao:2018ekp,CentellesChulia:2018gwr,CentellesChulia:2018bkz,Han:2018zcn,Borah:2018gjk,Borah:2018nvu,Calle:2018ovc,Carvajal:2018ohk,Ma:2019yfo,Saad:2019bqf,Dasgupta:2019rmf,
Enomoto:2019mzl,Jana:2019mez,Ma:2019iwj,Ma:2019byo,Restrepo:2019soi,CentellesChulia:2019xky,Calle:2019mxn,Kumar:2025cte,Borboruah:2024lli,De:2021crr,Luo:2020sho,Luo:2020fdt,CentellesChulia:2024iom}. 
To generate the tiny Dirac neutrino masses, an attractive approach is to attribute them to a radiative mechanism realized through loop diagrams. The loop suppression naturally explains the smallness of the masses, while the particles running in the loop can carry quantum numbers under a discrete symmetry. This same symmetry can stabilize the lightest particle in the loop, making it a viable dark matter (DM) candidate. 

The weakly interacting massive particle (WIMP) paradigm, featuring electroweak-scale particles with weak cross sections ($\sigma \sim10^{-26}$ cm$^3$/s), naturally yields the observed DM density ($\Omega_{\rm DM}h^2\approx0.12$) via thermal freeze-out (see~\cite{Arcadi:2017kky} for a review). However, this paradigm faces significant challenges from null results in DM direct detection~\cite{PandaX:2024qfu,LZ:2024zvo}. In response, the feebly interacting massive particle (FIMP) scenario has emerged as a compelling alternative~\cite{Hall:2009bx,Liu:2022jdq,Bian:2018bxr,Hessler:2016kwm,Molinaro:2014lfa}. Unlike thermal WIMPs, FIMPs are produced via extremely weak couplings, avoiding current experimental bounds while naturally explaining the observed dark matter abundance.

The Scotogenic Dirac Model, originally proposed in Ref.~\cite{Farzan:2012sa} and later developed with an alternative symmetry realization in Ref.~\cite{Guo:2020qin}, extends the SM with right-handed neutrinos $\nu_R$, vector-like fermions $N$, a scalar doublet $\Phi$, and a scalar singlet $\chi$. 
Previous work in Ref.~\cite{Guo:2020qin} explored the scenario where the lightest fermion $N_1$ acts as a WIMP dark matter. This scenario faces significant challenges due to stringent constraints from lepton flavor violation (LFV) experiments, which severely limit the Yukawa couplings required for efficient thermal relic production. Consequently, achieving the observed relic density requires substantial tuning of parameters and relies on annihilation channels mediated by the $y_\chi$ coupling.
In this work, we propose a novel realization of the Scotogenic Dirac Model in which the lightest neutral fremion $N_1$ serves as a FIMP DM candidate. This scenario is defined by the extreme suppression of the Yukawa couplings $(y_\Phi)_{\alpha 1}$ and $(y_\chi)_{\alpha 1}$, which ensures that $N_1$ never enters thermal equilibrium with the Standard Model (SM) plasma throughout cosmic history. This FIMP framework not only provides a natural mechanism for DM production but also has a profound impact on neutrino physics, leading to a rank-2 structure for the neutrino mass matrix. This structure predicts one nearly massless neutrino state, which is allowed by the oscillation data.
We systematically investigate two complementary production mechanisms for $N_1$. The first is the conventional freeze-in mechanism, where $N_1$ is slowly produced via the extremely small Yukawa couplings through decays of heavier $Z_2$-odd scalars and fermions. The second is the super-WIMP mechanism, where $N_1$ is produced by the late decays of a thermally frozen-out next-to-lightest odd particle (NLOP). The identity of the NLOP—whether it is $N_2$, $\phi_{R,I}$, $\phi^\pm$, or $\chi$—plays a crucial role in shaping the model's phenomenology, directly influencing the DM relic abundance and cosmological observables.
A central focus of our analysis is the effective number of relativistic species, $\Delta N_{\mathrm{eff}}$. We show that the deviation $\Delta N_{\mathrm{eff}}$ receives contributions from two distinct sources: a thermal component from the primordial bath of right-handed neutrinos, and a non-thermal component arising from energy injection due to late NLOP decays. Our comprehensive parameter scan demonstrates that viable regions of parameter space exist for all possible NLOP candidates, satisfying all current observational constraints from DM relic density, lepton flavor violation, BBN, CMB, and $\Delta N_{\mathrm{eff}}$.

The remainder of this paper is organized as follows: Section~\ref{sec:model} introduces the FIMP Scotogenic Dirac Model and the mechanism for Dirac neutrino mass generation. Section~\ref{sec:relic} details the dark matter relic density calculations. Section~\ref{sec:neff} analyzes the cosmological implications for the effective number of relativistic species. Finally, we make our conclusion in Section~\ref{sec:conclusion}.

\section{THE MODEL}
\label{sec:model}

The Scotogenic Dirac Model was originally proposed in Ref.~\cite{Farzan:2012sa}, and a different symmetry realization was put forward in~\cite{Guo:2020qin} by one of the authors. In Tab.~\ref{tab:particles} we list out the relevant particles in this model.
\begin{table}[h]
    \centering
    \caption{Particles and symmetries in the scotogenic Dirac model.}
    \renewcommand{\arraystretch}{1.2}
    \begin{tabular}{c|c|c||c|c|c|c}
        \hline
         & ~$F_L$~ & ~$H$~ & ~$\nu_R$~ & ~$N$~ & ~$\Phi$~ & ~$\chi$~ \\
        \hline
        $SU(2)_L$ & 2 & 2 & 1 & 1 & 2 & 1 \\
        $U(1)_Y$ & $-1/2$ & $1/2$ & 0 & 0 & $1/2$ & 0 \\
        \hline
        $Z_3$ & 0 & 0 & $\omega$ & $\omega$ & $\omega$ & 0 \\
        $Z_2$ & + & + & + & - & - & - \\
        \hline
    \end{tabular}
    \label{tab:particles}
\end{table}
Apart from the Standard Model lepton doublet $F_L = \left(\nu_L, \ell_L\right)^T$ and Higgs $H$, additional species are the right-handed neutrinos $\nu_R$, three vector-like fermions $N$, an extra scalar doublet $\Phi\equiv \left(\phi^+, (\phi_R + i\phi_I)/\sqrt{2}\right)^T$ and scalar singlet $\chi$. Tree-level Dirac neutrino masses and possible Majorana masses of $\nu_R$ and $N$ are forbidden due to the existence of $Z_3$. 

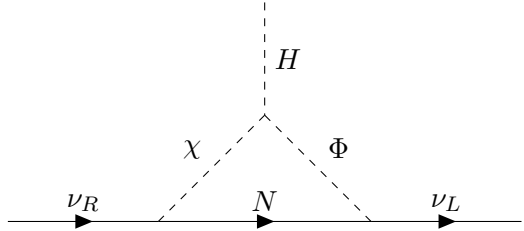
\begin{figure}
    \centering
    \begin{tikzpicture}
    \begin{feynman}
    \vertex (a); 
    \vertex [below=1.5cm of a] (b); 
    \vertex [below left=2cm of b] (c); 
    \vertex [below right=2cm of b] (d); 
    \vertex [left=2cm of c] (e); 
    \vertex [right=2cm of d] (f); 

    \diagram* {
      (a) -- [scalar, edge label=\(H\)] (b),
      (b) -- [scalar, edge label=\(\Phi\)] (d),
      (c) -- [scalar, edge label=\(\chi\)] (b),
      (e) -- [fermion, edge label=\(\nu_R\)] (c),
      (d) -- [fermion, edge label=\(\nu_L\)] (f),
      (c) -- [fermion, edge label=\(N\)] (d), 
    };
  \end{feynman}
    \end{tikzpicture}
    \caption{The one-loop neutrino mass in Scotogenic Dirac Model.}
    \label{fig:mass}
\end{figure}

The introduction of new particles in Tab.~\ref{tab:particles} would bring in the following interactions:
\begin{equation}
    - \mathcal{L}_{\rm new} \supset \left(y_\Phi \overline{F_L} \tilde \Phi N + y_\chi \overline{\nu_R} \chi N + {\rm h.c.} \right) + m_N \overline{N} N,
    \label{eq:yukawa}
\end{equation}
where $\tilde \Phi \equiv i \sigma_2 \Phi^\star$. Relevant scalar potential terms are given by
\begin{align}
    V = & -\mu_H^2 H^\dagger H + \mu_\Phi^2 \Phi^\dagger \Phi + \frac{1}{2} \mu_\chi^2 \chi^2 + \frac{1}{2} \lambda_1 (H^\dagger H)^2 + \frac{1}{2} \lambda_2 (\Phi^\dagger \Phi)^2 + \frac{1}{4!} \lambda_3 \chi^4 \notag \\
    & + \lambda_4 (H^\dagger H)(\Phi^\dagger \Phi) + \frac{1}{2} \lambda_5 (H^\dagger H) \chi^2 + \frac{1}{2} \lambda_6 (\Phi^\dagger \Phi) \chi^2 + \lambda_7 (H^\dagger \Phi)(\Phi^\dagger H) \notag \\
    & + \left( \mu \Phi^\dagger H \chi + \text{h.c.} \right).
    \label{eq:potential}
\end{align}
The scalar singlet $\chi$ is set as real, for simplicity. The $\mu$-terms in Eq.~\eqref{eq:potential} is the source which softly break the $Z_3$ symmetry, hence it's natural to keep it a small value. The one-loop neutrino mass (as illustrated in Fig.~\ref{fig:mass}) is then generated as
\begin{equation}
    (M_\nu)_{\alpha\beta} 
    = \frac{\sin 2\theta}{32\pi^2\sqrt{2}} \sum_{k=1,2,3} (y_\Phi)_{\alpha k} (y_\chi^*)_{\beta k} m_{N_k} \left( \frac{m_1^2}{m_1^2 - m_{N_k}^2} \ln \frac{m_1^2}{m_{N_k}^2} - \frac{m_2^2}{m_2^2 - m_{N_k}^2} \ln \frac{m_2^2}{m_{N_k}^2} \right).
    \label{eq:numass}
\end{equation}
Lower indices $\alpha(\beta)$ and $k$ stand for different generations of leptons and $N$s. The angle $\theta$ in Eq.~\ref{eq:numass} originates from mixing between the real component of scalar doublet $\Phi$ and singlet $\chi$. When the SM Higgs $H$ acquires its vacuum expectation value $v$ after electroweak spontaneous symmetry breaking, the angle is quantitatively fixed as
\begin{equation}
    \tan 2\theta = \frac{2\sqrt{2} \mu v}{\mu_\Phi^2 - \mu_\chi^2 + \left(\lambda_4 -\lambda_5\right) v^2}.
    \label{eq:mixingtheta}
\end{equation}
The $m^2_{1(2)}$ in Eq.~\ref{eq:numass} indicate eigenvalues of the $\chi-\phi_R$ mixing mass-square matrix
\begin{equation}
m_{(\chi,\phi_R)}^2 = \begin{pmatrix}
\mu_\chi^2 + \lambda_5 v^2 & \sqrt{2} \mu v \\
\sqrt{2} \mu v & \mu_\Phi^2 + (\lambda_4 + \lambda_7) v^2
\end{pmatrix}.
\end{equation}
Treating $\mu$ as a tiny parameter results in small mixing, justifying the approximation $m_{1(2)} \approx m_{\chi(\phi_R)}$. Hence we will use $m_{\chi}$ and $m_{\phi_R}$ hereafter.
The masses of the other scalars are $m_h^2 = 2 \lambda_1 v^2, ~m_{\phi^\pm}^2 = \mu_\Phi^2 + \lambda_4 v^2, ~ 
m_{\phi_I}^2 = \mu_\Phi^2 + (\lambda_4 + \lambda_7) v^2$. Here $h$ is the observed 125~GeV boson.

The assignment of $Z_2$ in Tab.~\ref{tab:particles} would make particles in loop of Fig.~\ref{fig:mass} to be in dark-sector, and the lightest one is naturally served as the dark matter particle candidate. In Ref.~\cite{Guo:2020qin}, we had made a detailed study on the situation that $N_1$ is a WIMP-type dark matter particle. 
In this work, however, we explore an alternative scenario where the dark matter candidate $N_1$ is a feebly interacting massive particle.
The FIMP nature of $N_1$ implies that the Yukawa couplings $(y_\Phi)_{\alpha 1}$ and $(y_\chi)_{\alpha 1}$ are extremely small. This setup significantly distinguishes the model from the conventional WIMP Dirac scotogenic framework. 
From Eq.~\ref{eq:numass}, we observe that the neutrino mass matrix $M_\nu$ is approximately of rank-2, a direct consequence of the suppressed couplings $(y_\Phi)_{\alpha 1}$ and $(y_\chi)_{\alpha 1}$. This structure predicts that one of the three neutrino mass eigenstates is nearly massless, which remains consistent with the oscillation data requiring at least two massive neutrinos.

The Yukawa interactions $y_\Phi$ could mediate processes leading to LFV, which is extremely suppressed in SM. On the experimental side, the most stringent limits arise from the rare decays $\ell_\alpha \to \ell_\beta \gamma$. The decay branching ratio is calculated as
\begin{equation}
    \text{Br}(\ell_\alpha \to \ell_\beta \gamma) = 
\text{Br}(\ell_\alpha \to \ell_\beta \nu_\alpha \overline{\nu}_\beta) \times 
\frac{3 \alpha_{\rm em}}{16 \pi G_F^2} 
\left| \sum_i \frac{(y_\Phi)_{\beta i} (y_\Phi^\ast)_{\alpha i}}{m_{\phi^\pm}^2} j \left( \frac{m_{N_i}^2}{m_{\phi^\pm}^2} \right) \right|^2,
\end{equation}
here $j(r) = \left(1-6r + 3r^2 + 2r^3 -6r^2\ln r\right)/\left(12(1-r)^4\right)$.
The latest result from MEG II reported the limit as $\text{Br}(\mu \to e \gamma) \lesssim 1.5 \times 10^{-13}$~\cite{MEGII:2025gzr}. This could be translated into limit on $y_\Phi$ as $(y_\Phi)_{\alpha 2,3} \lesssim 0.01 $, if we assume the new masses are not far away from the electroweak scale. When combined with the neutrino mass in Eq.~\ref{eq:numass}, and noting that the mixing angle $\theta$ is small (as it reflects the $Z_3$-breaking parameter $\mu$ in Eq.~\ref{eq:mixingtheta}), we deduce that $(y_\chi)_{\alpha 2,3}$ must be comparable to or larger than $(y_\Phi)_{\alpha 2,3}$. For example, with $\sin2\theta \sim 10^{-5}$ and $(y_\Phi)_{\alpha 2,3} \lesssim 0.01$, reproducing the observed neutrino masses requires $(y_\chi)_{\alpha 2,3} \gtrsim 0.01$.

\section{Dark Matter RELIC DENSITY}
\label{sec:relic}

The residual $Z_2$ symmetry confines loop particles to the dark sector, with the lightest $Z_2$-odd particle serving as dark matter candidate.
In~\cite{Guo:2020qin}, we analyzed the scenario where the fermion $N_1$ acts as WIMP dark matter. We found that the dark matter annihilation via $y_\Phi$ is difficult to achieve at the correct relic density. This limitation arises because the Yukawa coupling $y_\Phi$ faces stringent constraints from LFV. The $y_\Phi$ compatible with these LFV constraints substantially suppresses $N_1$ annihilation cross-sections, resulting in dark matter overproduction that is excluded by cosmological observations. Consequently, the dominant annihilation must proceed through $y_\chi$.

In this work, however, we would try a FIMP realization of dark matter $N_1$. The feeble interaction strength would retain $N_1$ always out of equilibrium. Thus the key ways to produce $N_1$ are decays of heavier $Z_2$-odd particles. The dominant contributions are from decays of scalars:
\begin{align}
\Gamma_{\phi_{R,I} \to N_1 \bar{\nu}_\alpha} &= \frac{\left | (y_{\Phi})_{\alpha 1} \right |^2}{32 \pi} \frac{ (m_{\phi_{R,I}}^2 - m_{N_1}^2)^2}{m_{\phi_{R,I}}^3}, \label{eq:phidecay1}\\
\Gamma_{\phi^{+} \to N_1 \ell^+_\alpha} &= \frac{\left| (y_{\Phi})_{\alpha 1}\right | ^2}{16 \pi} \frac{(m_{\phi^+}^2 - m_{N_1}^2)^2}{m_{\phi^+}^3}, \label{eq:phidecay2}\\
\Gamma_{\chi \to N_1 \bar{\nu}_\alpha} &= \frac{\left| (y_{\chi})_{\alpha 1} \right |^2}{16 \pi} \frac{(m_{\chi}^2 - m_{N_1}^2)^2}{m_{\chi}^3}. \label{eq:chidecay}
\end{align}
Additional production channels involve decays from $N_{2,3}$:
\begin{align}
\Gamma_{N_{2,3} \to N_1 \ell^-_\alpha \ell^+_\beta} &= \frac{ \left| (y_{\Phi})_{\beta 1}\right |^{2} \left| (y_{\Phi})_{\alpha 2,3}\right |^{2} }{6144 \pi^3} \frac{ m_{N_{2,3}}^3 ( m_{N_{2,3}}^2-  2m_{N_1}^2 )}{m_{\phi^+}^4}, \label{eq: Ndecay1}\\
\Gamma_{N_{2,3} \to N_1 \nu_\alpha \bar{\nu}_\beta} &= \frac{m_{N_{2,3}}^3 ( m_{N_{2,3}}^2-  2m_{N_1}^2 )}{6144 \pi^3} \left(\sum_{\phi_R,\phi_I} \frac{ \left| (y_{\Phi})_{\beta 1}\right |^{2} \left| (y_{\Phi})_{\alpha 2,3}\right |^{2} }{m_{\phi}^4} + \frac{ \left| (y_{\chi})_{\beta 1}\right |^{2} \left| (y_{\chi})_{\alpha 2,3}\right |^{2} }{m_\chi^4} \right) . \label{eq:Ndecay2}
\end{align}
Fermionic decay contributions remain subdominant due to both three-body phase space suppression and additional Yukawa coupling factors. Moreover, decays into $\bar{N}_1$ and decays of $\bar{N}_2 \to N_1(\bar{N}_1)$, whose widths are similar as in Eqns.~\ref{eq:phidecay1} to \ref{eq:Ndecay2}, should also be taken into account to get the total abundance of dark matter. To ensure $N_1$ remains out of thermal equilibrium, the decay rate should be smaller than the Hubble expansion rate. A rough estimate requires the Yukawa coupling $(y_{\Phi,\chi})_{\alpha 1} \lesssim 10^{-7}$, for scalar masses at $\mathcal{O}(\rm TeV)$.

The evolution of dark matter and relevant particles are governed by the following Boltzmann equations:
\begin{align}
    \frac{dY_{N_1+\bar{N}_1}}{dz} &= 
    k^{*} z \sum_{X} \left(\tilde{\Gamma}_{X \to N_1} + \tilde{\Gamma}_{X \to \bar N_1}\right) Y_X ,\\
    \frac{dY_X}{dz} &= k^{*} z \sum_{A} \tilde{\Gamma}_{A \to X} \left( Y_A - \frac{Y_A^{\text{eq}}}{Y_X^{\text{eq}}} Y_X \right) - k^{*} z \sum_{B} \tilde{\Gamma}_{X \to B} \left( Y_X - \frac{Y_X^{\text{eq}}}{Y_B^{\text{eq}}} Y_B \right) \notag \\
    &+ \frac{k}{z^2} \langle \sigma v\rangle_{{\rm SM}\to X \bar X} \left( (Y_{\rm SM}^{\rm eq})^2 - \frac{(Y_{\rm SM}^{\rm eq})^2}{(Y_X^{\text{eq}})^2} Y_X Y_{\bar X} \right).
\end{align}
Here $ X = \phi_R, \phi_I, \phi^\pm , \chi, N_{2,3}(\bar N_{2,3})$ stands for all the $Z_2$-odd particles that could decay into dark matter. For a specific $X$, its evolution is affected by three kinds of processes: decays of heavier $A$ into $X$, decays of $X$ into lighter $B$, and SM pairs annihilating into $X$ pairs. 
The dimensionless parameter $z$ is defined as the ratio of the dark matter mass to the evolution temperature, i.e. $z \equiv m_{N_1} / T$. The parameter $k^{*}$ is given by $k^{*}=\sqrt{45/(4 \pi^3 g^*)} M_{\rm Pl} /m_{N_1}^{2}$, $g^*$ is the effective number of degrees of freedom of the relativistic particles, $ M_{{\rm Pl}} = 1.2 \times 10^{19}~\GeV$ is the Planck mass. Thermal decay width $\tilde{\Gamma}_{i\to j}$ is defined as $\tilde{\Gamma } _{i\to j} =\Gamma_{i\to j} K_1/K_2$, where $K_1$ and $K_2$ denote the first and second order of modified Bessel functions of the second kind, respectively. $Y_i^{\text{eq}}$ represents the abundance in equilibrium for species $i$.

\begin{figure}
    \centering 
    \includegraphics[width=0.5\linewidth]{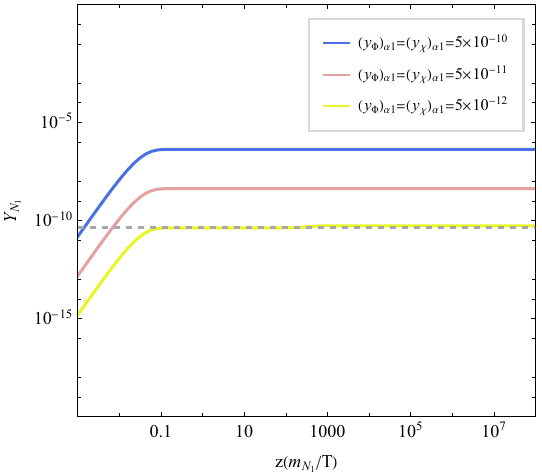}
    \caption{The evolution of dark matter through freeze-in. The parameters used are listed in Tab.~\ref{tab:params}.}
    \label{fig:freezein}
\end{figure}

In practice, we write down the model using $\texttt{FeynRules}$~\cite{Alloul:2013bka} and the output model files are used to calculate relic density by $\texttt{micrOmegas}$~\cite{Alguero:2023zol}.
We present the results of this calculation in Fig.~\ref{fig:freezein}, which illustrates the comoving yield of $N_1$ as a function of the cosmological variable $z = m_{N_1}/T$ for several benchmark values of the feeble Yukawa couplings $(y_\Phi)_{\alpha 1}$ and $(y_\chi)_{\alpha 1}$. For simplicity in this analysis, we set these couplings to be equal, i.e., $(y_\Phi)_{\alpha 1}=(y_\chi)_{\alpha 1}$, while the remaining model parameters are fixed to the values specified in Tab.~\ref{tab:params}.
The resulting evolution curves exhibit the characteristic hallmark of the freeze-in mechanism. At high temperatures ($z \ll 1$), the abundance of $N_1$ is negligible. As the universe cools, $N_1$ is slowly but steadily produced from decays of the heavier $Z_2$-odd particles. The yield grows progressively until the source particles become thermally suppressed and depleted, at which point the comoving number density of $N_1$ ``freezes in" and remains constant. For reference, the observed dark matter relic density, $\Omega_{\rm DM} h^2 \approx 0.12$, is indicated by a horizontal dashed line. We demonstrate a viable parameter point, corresponding to the specific coupling strength $(y_\Phi)_{\alpha 1} = (y_\chi)_{\alpha 1} = 5 \times 10^{-12}$, successfully saturates the observed dark matter density.

\begin{table}
\centering
\caption{Parameter sets used in Fig.~\ref{fig:freezein} and Fig.~\ref{fig:NLOP}, masses are given in GeV.}
\renewcommand{\arraystretch}{1.2}
\begin{tabular}{c c c c c c| c c c c|c}
\hline
  $m_{N_1}$ & $m_{N_2}$ &  $m_{N_3}$ & $m_{\phi_R}$ & $m_{\phi_I, \phi^\pm}$  & $m_\chi$ & $(y_{\Phi,\chi})_{\alpha 1}$ & $(y_{\Phi})_{\alpha 2,3}$ & $(y_{\chi})_{\alpha 2,3}$ & $\lambda_5$ & Figure \\
\hline
  $10$ & $800$ & $800$ & $800$ & $800$ & $800$ & $5\times10^{-10,-11,-12}$ & $10^{-3}$ & $10^{-2}$  & $10^{-3}$ & Fig.~\ref{fig:freezein}\\
  \hline
  $10$ & $600/770$ & $800$ & $800$ & $800$ & $800$ & $5\times10^{-13}$ & $10^{-2}$ & $10^{-2}$ & $10^{-3}$ & \cref{subfig:N2coan}\\
  $10$ & $300$ & $800$ & $800$ & $800$ & $800$ & $5\times10^{-13}$ & $10^{-2}$ & $0.1/0.3$ & $10^{-3}$ & \cref{subfig:N2ychi}\\
  $500$ & $800$ & $800$ & $750$ & $800$ & $800$ & $5\times10^{-13}$ & $10^{-3}$ & $10^{-2}$ & $10^{-3}$ & \cref{subfig:phiR}\\
  $10$ & $800$ & $800$ & $800$ & $800$ & $550$ & $5\times10^{-13}$ & $10^{-3}$ & $0.5/10^{-3}$ & $10^{-3}/0.02$ & \cref{subfig:chi}\\
\hline
\end{tabular}
\label{tab:params}
\end{table}

\begin{figure}[htbp]
  \centering
  \begin{subfigure}[t]{0.45\textwidth}
    \caption{}
    \includegraphics[width=\textwidth]{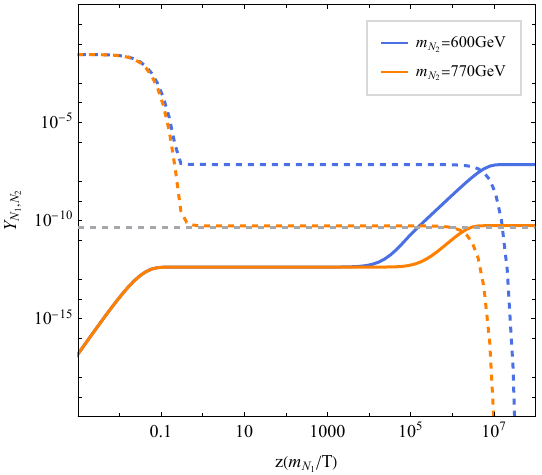}
    \label{subfig:N2coan}
  \end{subfigure}
  \begin{subfigure}[t]{0.45\textwidth}
    \caption{}
    \includegraphics[width=\textwidth]{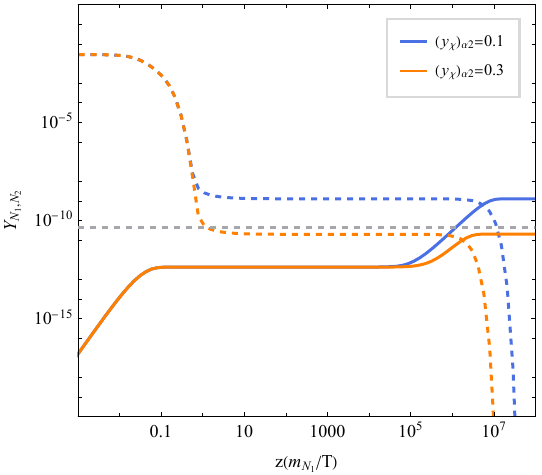}
    \label{subfig:N2ychi}
  \end{subfigure}
  \vspace{0.5cm} 
  \begin{subfigure}[t]{0.45\textwidth}
    \caption{}
    \includegraphics[width=\textwidth]{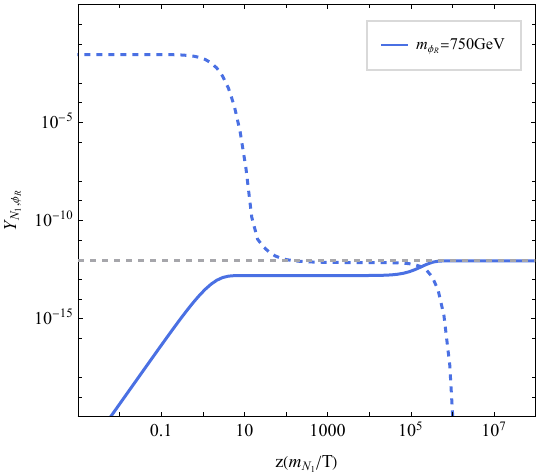}
    \label{subfig:phiR}
  \end{subfigure}
  \begin{subfigure}[t]{0.45\textwidth}
    \caption{}
    \includegraphics[width=\textwidth]{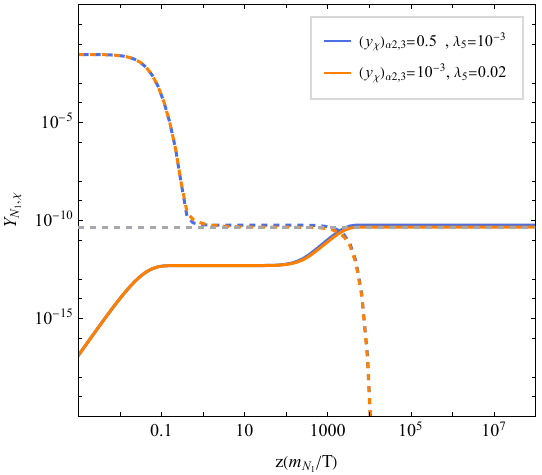}
    \label{subfig:chi}
  \end{subfigure}
  \caption{The evolution of dark matter and NLOP under the ``super-WIMP" mechanism. $N_2$ is chosen as NLOP in the upper two subfigures, while in the lower two subfigures $\phi_R$ and $\chi$ are selected as NLOP, respectively. The parameters used are listed in Tab.~\ref{tab:params}.}
  \label{fig:NLOP}
\end{figure}

In addition to the conventional freeze-in mechanism, dark matter can also be produced via the late decay of a thermally frozen-out next-to-lightest odd particle (NLOP), a scenario known as the ``super-WIMP'' mechanism~\cite{Feng:2003xh}. In this framework, heavier $Z_2$-odd particles decay predominantly into the NLOP in the early universe, as these transitions are not suppressed by small couplings. The NLOP is initially in thermal equilibrium with both SM and other $Z_2$-odd particles, but as the universe cools, its interaction rate drops below the Hubble expansion rate, leading to decoupling and the freeze-out of its comoving number density. Subsequently, the NLOP undergoes a highly suppressed decay into the stable dark matter, due to an extremely weak coupling. This results in a long-lived NLOP and a delayed, non-thermal production of dark matter that persists until the NLOP population fully decays.
The resulting dark matter relic abundance is totally fixed from the NLOP's freeze-out density:
\begin{equation}
    \Omega_{N_1}^{\text{super-WIMP}} = \frac{m_{N_1}}{m_{\text{NLOP}}} \Omega_{\text{NLOP}}^{\text{freeze-out}}.
\end{equation}
This production mechanism is illustrated in Fig.~\ref{fig:NLOP}, which showcases several distinct realizations of the NLOP within our model.

Figs.~\ref{subfig:N2coan} and~\ref{subfig:N2ychi} depict scenarios where the NLOP is the fermionic state $N_2$. Its freeze-out is driven by annihilations into SM leptons, mediated by the Yukawa couplings $(y_\Phi)_{\alpha 2}$ or $(y_\chi)_{\alpha 2}$. However, as shown in~\cite{Molinaro:2014lfa}, LFV constraints severely limit $(y_\Phi)_{\alpha 2}$, suppressing the annihilation cross section and leading to an overproduction of $N_2$ if this coupling dominates. To achieve the correct relic abundance, additional mechanisms are required.
One such mechanism is coannihilation: when $N_2$ is nearly degenerate in mass with other $Z_2$-odd states, their combined interactions enhance the effective annihilation rate. Fig.~\ref{subfig:N2coan} compares two benchmarks: $m_{N_2} = 600~\GeV$ and $770~\GeV$, with the other odd particles fixed at $800~\GeV$. The heavier $N_2$ case benefits from stronger coannihilation effects, yielding a reduced freeze-out abundance consistent with observations.
Alternatively, in our model, the coupling $(y_\chi)_{\alpha 2}$ opens an additional $t$-channel annihilation via exchange of the scalar singlet $\chi$, producing neutrino final states. As demonstrated in Fig.~\ref{subfig:N2ychi}, increasing $(y_\chi)_{\alpha 2}$ significantly enhances the annihilation rate, lowering $\Omega_{N_2}^{\text{freeze-out}}$ and bringing the final dark matter density into agreement with cosmological data.

In scenarios where the scalar doublet $\Phi$ serves as the NLOP, we focus on its neutral real component $\phi_R$ as the representative case (with $\phi_I$ and $\phi^\pm$ exhibiting similar dynamics). $\phi_R$ annihilates efficiently through electroweak gauge interactions (e.g., $\phi_R \phi_R \to W^+W^-, ZZ$), resulting in strong thermal coupling. The correct relic abundance is only achieved in mass region of $m_{\phi_R} > 500~\GeV$ (or $m_{\phi_R} < m_W$)~\cite{Barbieri:2006dq,LopezHonorez:2006gr,Cirelli:2005uq,Hambye:2009pw}.
In Fig.~\ref{subfig:phiR}, we consider a benchmark mass $m_{\phi_R} = 750~\GeV$. A moderate mass hierarchy between $N_1$ and $\phi_R$ is required (hence we set $m_{N_1} = 500~\GeV$), as the freeze-out abundance of $\phi_R$ is typically not significantly larger than the observed dark matter density \cite{Molinaro:2014lfa}. 

\begin{figure}
    \centering
    \includegraphics[width=0.5\linewidth]{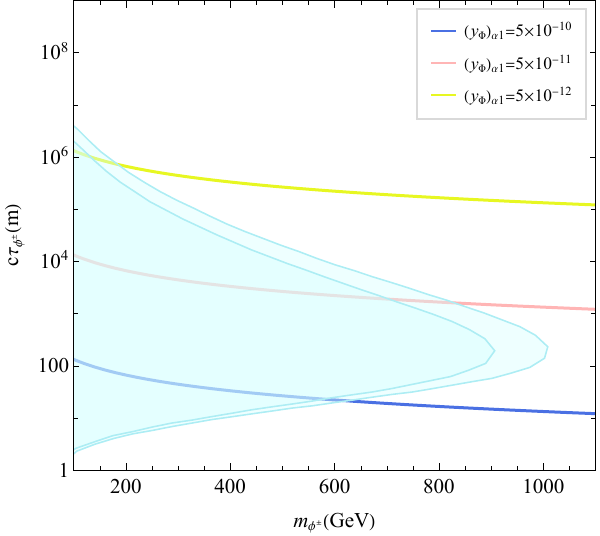}
    \caption{The projected detecting capabilities of MATHUSLA to a long-lived $\phi^\pm$, in the mass-decay length plane. The band shows the efficiencies vary from 0.5 to 1. Colored lines correspond to different strengths of Yukawa coupling between $\phi^\pm$ and the dark matter.}
    \label{fig:MAT}
\end{figure}

For completeness, we also consider the phenomenology of the charged component $\phi^\pm$ when it acts as the NLOP. Being electrically charged and long-lived (due to suppressed decays into $N_1$), $\phi^\pm$ behaves as a long-lived charged particle. The ATLAS Collaboration has recently searched for such states in $140~\mathrm{fb}^{-1}$ of $pp$ collision data at $\sqrt{s} = 13~\mathrm{TeV}$, using signatures based on high specific ionization energy loss and time-of-flight measurements~\cite{ATLAS:2025fdm}. Interpreting their results in the context of stau-to-gravitino decays, ATLAS excludes masses up to $560~\mathrm{GeV}$ for such long-lived charged particles.
Furthermore, in Fig.~\ref{fig:MAT} we give the assessment of the projected sensitivity of future long-lived particle detector MATHUSLA~\cite{Curtin:2018mvb}. Fig.~\ref{fig:MAT} shows the projected reach in the mass-decay length plane of a long-lived $\phi^\pm$~\cite{Curtin:2018mvb}, assumed to be pair-produced via Drell-Yan processes at the 14~TeV HL-LHC and to decay into dark matter. The reconstruction efficiency is assumed in 0.5--1. Contours indicate regions where at least four events are expected with 3~ab$^{-1}$. In Fig.~\ref{fig:MAT} we also show the Yukawa coupling strength as a function of mass and decay length. Nevertheless, for the parameter space that yields the correct dark matter relic density, MATHUSLA's projected coverage remains limited to only a marginal segment.

Finally, we consider the scenario in which the scalar singlet $\chi$ serves as the NLOP. 
The annihilation of $\chi$ in the early universe is governed by three key parameters: the Yukawa coupling $y_\chi$, the Higgs portal coupling $\lambda_5$, and the trilinear scalar coupling $\mu$. 
Annihilation through $y_\chi$ proceeds via $t$-channel exchange of the heavy neutral fermions $N_{2,3}$, yielding neutrino final states. 
Alternatively, $\chi$ can annihilate into SM particle pairs—such as $b\bar{b}$, $WW$, or $ZZ$—through $s$-channel Higgs exchange, mediated by the $\lambda_5 (H^\dagger H) \chi^2$ interaction.
Motivated by the requirement of naturally small neutrino masses and the preservation of an approximate global symmetry in the scalar sector, we treat the coupling $\mu$ as hierarchically suppressed throughout this analysis. 
This naturalness argument renders $\mu$ negligible for thermal processes, leaving $y_\chi$ and $\lambda_5$ as the dominant drivers of $\chi$'s annihilation dynamics.
To illustrate the viable parameter space, we examine two representative benchmark scenarios. 
The first features a large Yukawa coupling $(y_\chi)_{2,3} = 0.5$ with a small Higgs portal $\lambda_5 = 10^{-3}$, favoring annihilation into neutrinos. 
The second adopts a larger portal coupling $\lambda_5 = 0.02$ while keeping $(y_\chi)_{2,3} = 10^{-3}$ small, enhancing annihilation into SM states via Higgs mediation. 
As shown in Fig.~\ref{subfig:chi}, both configurations yield a $\chi$ freeze-out abundance that, after subsequent decay into $N_1$, reproduces the observed dark matter relic density. 
This demonstrates the flexibility of the singlet $\chi$ NLOP scenario in achieving the correct dark matter yield through distinct but equally viable annihilation pathways.

\section{Constraints from $N_{\text{eﬀ}}$}
\label{sec:neff}

The Dirac nature of neutrinos necessitates both left-handed ($\nu_L$) and right-handed ($\nu_R$) chiral components. The existence of $\nu_R$ contributes additional relativistic degrees of freedom, thereby increasing the radiation energy density in the early universe. This contribution is commonly parameterized by the effective number of relativistic species, defined as
\begin{equation}
    N_{\rm eff} = \frac{8}{7} \left(\frac{11}{4}\right)^{4/3} \frac{\rho_{\nu_L} + \rho_{\nu_R}}{\rho_\gamma} = 3 \left(\frac{11}{4}\right)^{4/3} \left[ \left(\frac{T_{\nu_L}}{T_{\gamma}}\right)^4 + \left(\frac{T_{\nu_R}}{T_{\gamma}}\right)^4 \right].
    \label{eq:Neff}
\end{equation}
Within the SM, the effective number of relativistic species is precisely calculated to be $N_{\rm eff}^{\rm SM} = 3.045$~\cite{Mangano:2005cc,Grohs:2015tfy,deSalas:2016ztq}, incorporating effects from neutrino oscillations, non-thermal spectral distortions, and finite-temperature corrections. Any deviation from this value due to physics beyond the Standard Model (BSM) is typically expressed as $\Delta N_{\rm eff} \equiv N_{\rm eff} - N_{\rm eff}^{\rm SM}$. 
The Planck2018 results provide the most accurate and stringent constraint to date, yielding $\Delta N_{\rm eff} \leq 0.285$ at the $2\sigma$ confidence level~\cite{Planck:2018vyg}. While the upcoming CMB-S4 experiment is projected to significantly improve sensitivity, with a forecasted reach of $\Delta N_{\rm eff} \leq 0.06$ at the same confidence level~\cite{CMB-S4:2016ple}.

\subsection{Thermal Contribution}
\begin{figure}
    \centering
    \includegraphics[width=0.45\linewidth]{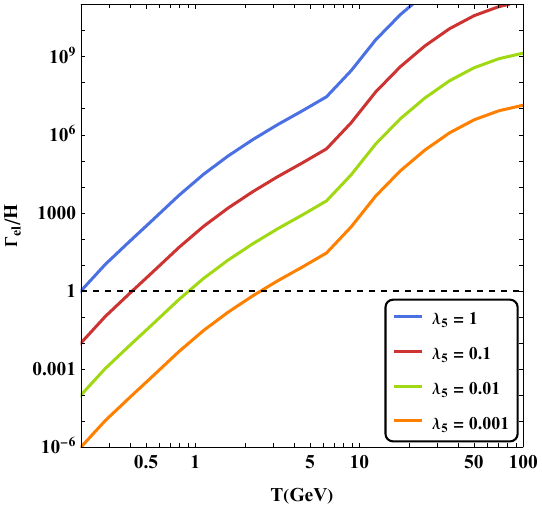}
    \includegraphics[width=0.44\linewidth]{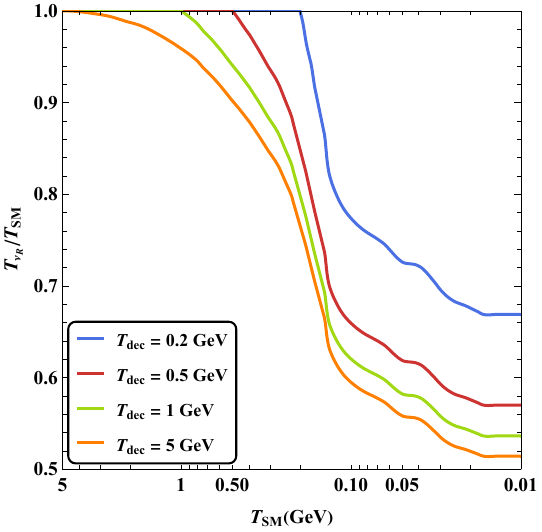}
    \caption{(Left panel) The ratio of $\Gamma_{\rm el}/H$ as a function of temperature for different choice of $\lambda_5$. Here we have chosen $m_\chi = 100~\GeV$ for demonstration. (Right panel) The evolution of ratio $T_{\nu_R}/T_\gamma$ for different dark plasma decoupling temperature.}
    \label{fig:gamma2H}
\end{figure}

In the early hot universe, all particles except the FIMP $N_1$ were in thermal equilibrium with the SM plasma. As the universe expanded and cooled, the dark plasma---comprising the scalar singlet $\chi$, fermion singlets $N_{2,3}$, and right-handed neutrinos $\nu_R$---gradually decoupled from the SM plasma. However, partial thermal equilibrium was maintained within the dark plasma due to the large Yukawa coupling $y_\chi$.
Interactions between the dark plasma and the SM plasma can occur through either $N_{2,3}$ or $\chi$. For $N_{2,3}$, interactions with the SM are mediated by the coupling $y_\Phi$, with the scattering amplitude scaling as $|y_\Phi|^2$. In contrast, $\chi$ couples to the SM either via $t$-channel Higgs exchange or through a contact interaction with Higgs, with the scattering amplitude proportional to $\lambda_5$.
Due to stringent constraints from LFV, $y_\Phi$ is typically too small to support significant interactions. As a result, the dominant portal between the dark plasma and the SM is mediated by $\chi$. When $\chi$ eventually decouples from the SM plasma, the temperature of the right-handed neutrinos $\nu_R$ begins to deviate from that of the SM thermal bath.
The decoupling temperature $T_{\text{dec}}$ is determined by the condition:
$\Gamma_{\text{el}}(T_{\text{dec}}) = H(T_{\text{dec}})$,
where $\Gamma_{\text{el}} \equiv \sum n_{\text{SM}} \langle \sigma v \rangle_{\chi\,\text{SM} \to \chi\,\text{SM}} / (m_\chi / T)$ denotes the effective elastic scattering rate between $\chi$ and SM particles, accounting for the number of scatterings required to transfer energy of order $T$.
In Fig.~\ref{fig:gamma2H}, we show the ratio $\Gamma_{\rm el}/H$ as a function of temperature for different values of $\lambda_5$. The dashed line, corresponding to $\Gamma_{\rm el}/H = 1$, indicates the condition for decoupling.
The temperature ratio $T_{\nu_R}/T_{\rm SM}$ can be determined using entropy conservation:
\begin{equation}
    \frac{T_{\nu_R}}{T_{\rm SM}} = \left( \left. \frac{g_{\rm DP}^{*s}}{g_{\rm SM}^{*s}} \right|_{T_{\rm dec}} \frac{g_{\rm SM}^{*s}}{g_{\rm DP}^{*s}} \right)^{1/3},
    \label{eq:Tratio}
\end{equation}
where $g_X^{*s}$ denotes the effective number of relativistic degrees of freedom, with respect to entropy density, in the $X$ plasma (with ``DP'' referring to the dark plasma). The evaluation at $T_{\rm dec}$ indicates that the ratio is set at the dark plasma decoupling temperature. The evolution of $T_{\nu_R}/T_{\rm SM}$ is shown in Fig.~\ref{fig:gamma2H} for different choices of the dark plasma decoupling temperature. The later the dark plasma decouples (which corresponds to a larger $\lambda_5$, and $\nu_R$ is heated more efficiently by the SM plasma over a longer period), the higher temperature ratio we will arrive at. This finally results in a larger contribution to the effective number of relativistic species. 

\subsection{Non-thermal Contribution from Delayed Decay of NLOP}
In addition to thermal contributions from $\nu_R$, the delayed decay of the NLOP—particularly during the BBN or CMB epochs—can leave observable imprints on cosmological data. 
Several studies in the literature have examined the late decays of heavy relics into electron-positron pairs~\cite{Poulin:2016anj} or neutrinos~\cite{Hambye:2021moy}, and constraints have been derived based on their effects on BBN, CMB anisotropies and spectral distortions. In our model, the NLOP candidates include $N_2$, $\phi_{R,I}$, $\phi^\pm$, and $\chi$ (The roles of $\phi_R$ and $\phi_I$ as the NLOP are entirely analogous, therefore the following discussion for $\phi_R$ as the NLOP applies equally to $\phi_I$). Among these, $N_2$ and $\phi^\pm$ can decay into SM charged leptons. 
Exotic electromagnetic (EM) energy injection is tightly constrained by cosmology. CMB observations (e.g., Planck) limit such injection as it alters recombination and distorts temperature and polarization spectra, while COBE-FIRAS rules out spectral distortions from early injection. During BBN, excess energy disrupts light-element abundances, conflicting with observed D and $^4$He. Long-lived $N_2$ and $\phi^\pm$ with lifetimes between $3\times 10^4\,\mathrm{s}$ and recombination are strongly disfavored unless nearly degenerate with the dark matter (hence minimizing energy release)~\cite{Lucca:2019rxf,Poulin:2016anj,Kawasaki:2004qu,Jedamzik:2007qk}. In this work, we impose a conservative upper bound, requiring the lifetimes of $N_2$ and $\phi^\pm$ to be less than $3\times 10^4\,\mathrm{s}$.
For scenarios in which $\phi_{R}$ or $\chi$ is the NLOP, their delayed decays can inject energy into the left- or right-handed neutrinos, thereby altering the neutrino temperatures. The evolution of $T_{\nu_L}$, $T_{\nu_R}$, and $T_\gamma$ is governed by the following system of differential equations:
\begin{align}
\frac{d T_{\nu_L}}{d t} & = -H T_{\nu_L} + \frac{ \frac{\delta \rho_{\nu_{L}}}{\delta t} + \varepsilon^{\nu_L}_{\rm NLOP} \frac{\rho_{\rm NLOP}}{\tau_{\rm NLOP}} }{3 \frac{\partial \rho_{\nu_L}}{\partial T_{\nu_L}}}, \notag\\
\frac{d T_{\nu_R}}{d t} & = -H T_{\nu_R} + \frac{ \varepsilon^{\nu_R}_{\rm NLOP} \frac{\rho_{\rm NLOP}}{\tau_{\rm NLOP}} }{3 \frac{\partial \rho_{\nu_R}}{\partial T_{\nu_R}}}, \label{eq:Tode}\\
\frac{d T_{\gamma}}{d t} & = -\frac{4 H \rho_{\gamma} + 3 H(\rho_{e}+p_{e}) + \frac{\delta \rho_{\nu_{L}}}{\delta t}}{ \frac{\partial \rho_{\gamma}}{\partial T_{\gamma}} + \frac{\partial \rho_{e}}{\partial T_{\gamma}} }. \notag
\end{align}
Here, $\rho$ and $p$ denote energy and pressure densities, respectively. $\rho_{\rm NLOP}$ stands for the energy density of NLOP, it evolves as
\begin{equation}
    \frac{d\rho_{\rm NLOP}}{dt} = -3 H \rho_{\rm NLOP} - \frac{\rho_{\rm NLOP}}{\tau_{\rm NLOP}}.
    \label{eq:rhoNLOP}
\end{equation}
${\delta \rho}/{\delta t}$ in Eq.~\ref{eq:Tode} represents the energy transfer rate between the SM neutrinos and EM plasma due to weak interactions, as detailed in Refs.~\cite{Escudero:2018mvt, EscuderoAbenza:2020cmq}. The parameter $\varepsilon^{\nu_L(\nu_R)}_{\rm NLOP} = (m_{\rm NLOP}^2 - m_{N_1}^2)/(2m_{\rm NLOP}^2)$ quantifies the fraction of the decaying NLOP's rest energy that is effectively deposited into $\nu_L(\nu_R)$\footnote{A portion of the energy injected into $\nu_L$ is subsequently transferred to photons and $e^\pm$ via weak interactions, contributing to the thermalization of the EM plasma. The corresponding energy transfer fraction, denoted as $\xi_{\mathrm{EM}}$ in Ref.~\cite{Hambye:2021moy}, can be read from Figure~1 of that work. For NLOP masses around the electroweak scale, $\xi_{\mathrm{EM}}$ is highly suppressed and effectively negligible. Therefore, we neglect this energy transfer in our analysis.}.
The initial conditions for this system are set at a time when $\nu_L$ and the photon bath were in thermal equilibrium ($T_{\nu_L} = T_\gamma$), and the NLOP has not yet decayed. In practice, we begin the numerical calculation at $T_\gamma = 10~\mathrm{MeV}$, where $\nu_L$ is still coupled to the EM plasma, and any decays occurring earlier would contribute negligibly to the entropy of $\nu_L$ or $\nu_R$. The starting temperature of $\nu_R$ is fixed by Eq.~\ref{eq:Tratio} with setting $T_{\rm SM} = 10~\rm MeV$. Following Ref.~\cite{Hambye:2021moy}, we define a dimensionless parameter $f_{\rm NLOP} \equiv \Omega_{\rm NLOP} / \Omega_{\rm DM}$, where $\Omega_{\rm NLOP}$ represents the hypothetical present-day relic abundance of the NLOP if it were stable. Consequently, the initial energy density of the NLOP in Eq.~\ref{eq:rhoNLOP} can be expressed as a function of $f_{\rm NLOP}$.
In principle, a similar differential equation system should be solved for $N_2$ or $\phi^\pm$ as the NLOP. However, we have verified that the BBN and CMB constraints (i.e. $\tau < 3 \times 10^4\,\mathrm{s}$) ensure that any late-decay effects on $\Delta N_{\rm eff}$ are negligible.

\begin{figure}
    \centering
    \includegraphics[width=0.32\linewidth]{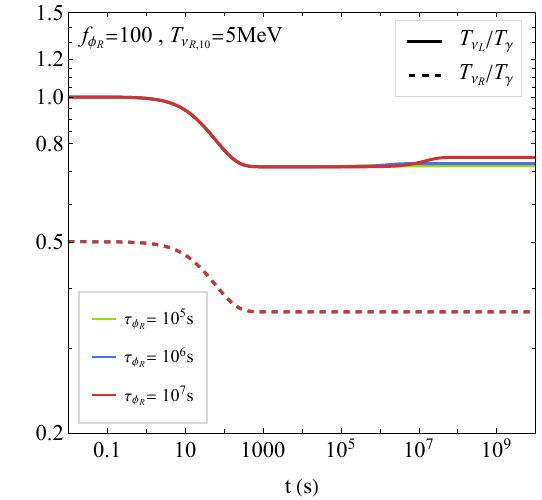}
    \includegraphics[width=0.32\linewidth]{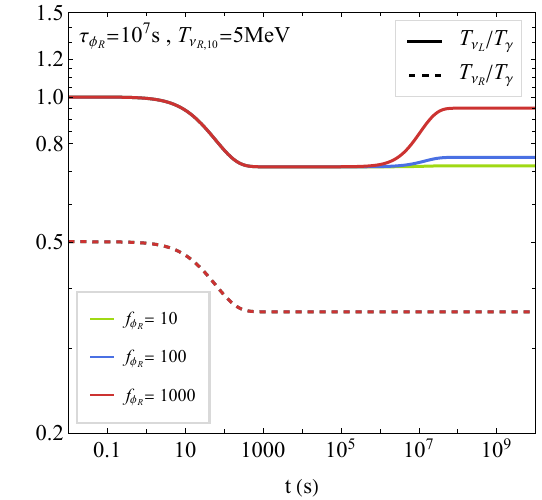}
    \includegraphics[width=0.32\linewidth]{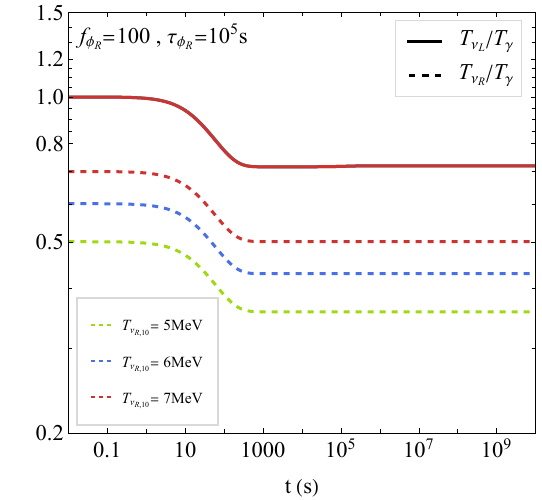}
    \includegraphics[width=0.32\linewidth]{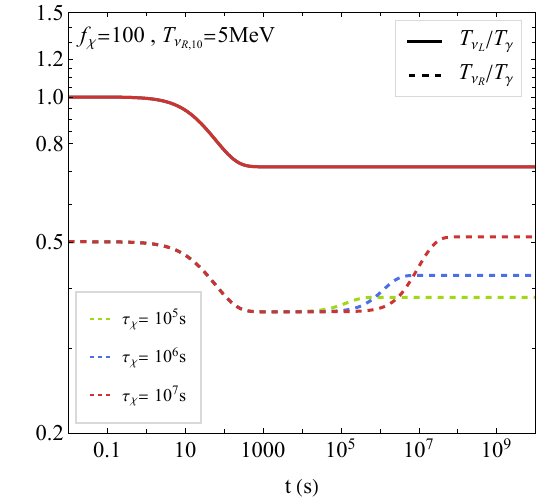}
    \includegraphics[width=0.32\linewidth]{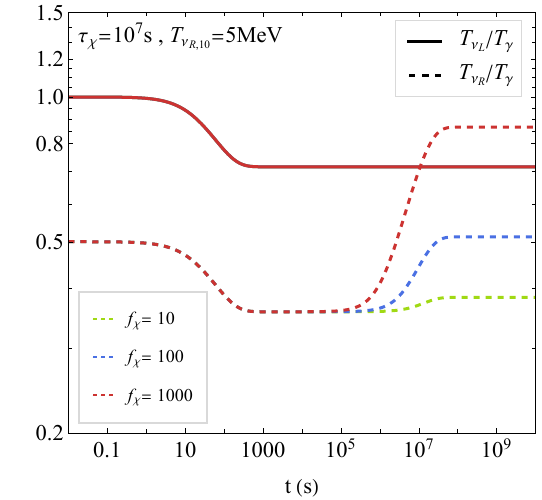}
    \includegraphics[width=0.32\linewidth]{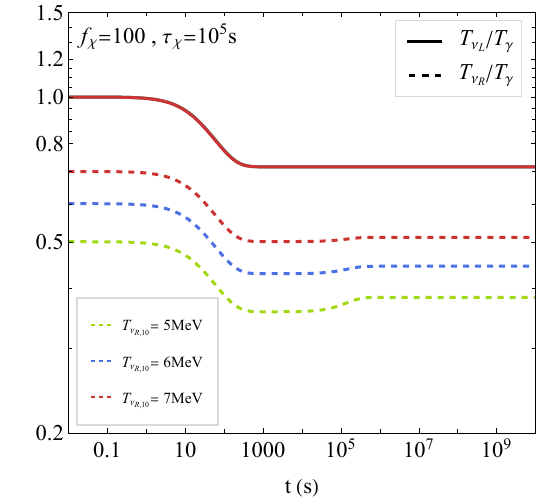}
    \caption{The evolution of the temperature ratios $T_{\nu_L}/T_\gamma$ (solid lines) and $T_{\nu_R}/T_\gamma$ (dashed lines). The upper panel shows the case where $\phi_R$ is the NLOP, and the lower panel shows the case where $\chi$ is the NLOP.}
    \label{fig:Tratio}
\end{figure}

We show the evolution of $T_{\nu_L}/T_\gamma$ (solid lines) and $T_{\nu_R}/T_\gamma$ (dashed lines) in Fig.~\ref{fig:Tratio}, where the first row corresponds to $\phi_R$ as the NLOP and the second row to $\chi$ as the NLOP. 
The input parameters for Eqs.~\eqref{eq:Tode} and \eqref{eq:rhoNLOP} are the NLOP energy fraction $f_{\mathrm{NLOP}}$, its lifetime $\tau_{\mathrm{NLOP}}$, and the temperature of the right-handed neutrino bath at $T_\gamma = 10~\mathrm{MeV}$, denoted $T_{\nu_R,10}$. 
We illustrate how different choices of these parameters influence the resulting temperature ratios.
In the first column, we fix $f_{\mathrm{NLOP}} = 100$ and $T_{\nu_R,10} = 5~\mathrm{MeV}$. 
The NLOP lifetimes are set to $10^5~\mathrm{s}$, $10^6~\mathrm{s}$, and $10^7~\mathrm{s}$ for illustration. 
As the lifetime of $\phi_R(\chi)$ increases, the delayed decay injects energy into $\nu_L(\nu_R)$ at a later cosmic time, leading to a more pronounced heating effect on $\nu_L(\nu_R)$.
In the second column, we fix $\tau_{\mathrm{NLOP}} = 10^7~\mathrm{s}$ and $T_{\nu_R,10} = 5~\mathrm{MeV}$, and vary $f_{\mathrm{NLOP}} = 10$, $100$, and $1000$. 
A larger $f_{\phi_R}(\chi)$ implies a higher relic density after freeze-out, resulting in greater energy injection into $\nu_L(\nu_R)$ once the decays are complete.
In the third column, we examine the impact of varying $T_{\nu_R,10}$ on the temperature ratio evolution. 
When $\phi_R$ is the NLOP, its late decay does not affect the thermal history of $\nu_R$, and hence there is no heating of $\nu_R$. 
When $\chi$ is the NLOP, variations in $T_{\nu_R,10}$ do not influence $T_{\nu_L}/T_\gamma$, since $\chi$ decays exclusively into $\nu_R$. 
A higher $T_{\nu_R,10}$ results in a larger $T_{\nu_R}/T_\gamma$ at the completion of $\chi$ decay.

\begin{figure}
    \centering
    \includegraphics[width=0.45\linewidth]{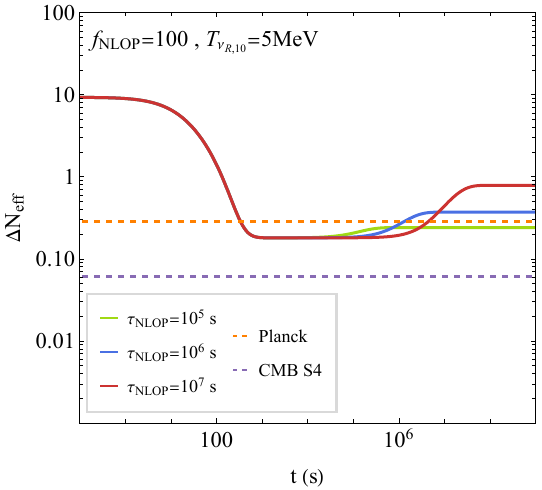}
    \includegraphics[width=0.45\linewidth]{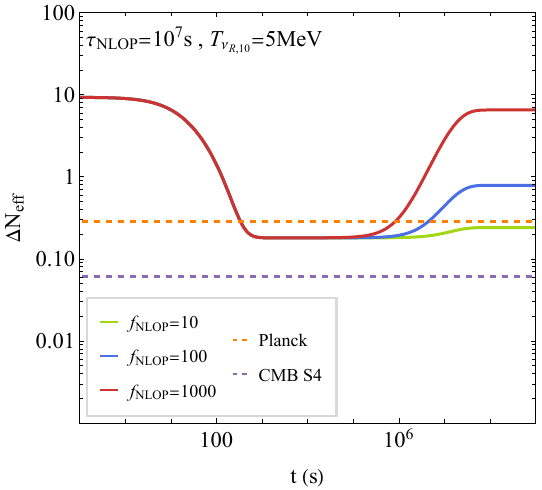}
    \includegraphics[width=0.45\linewidth]{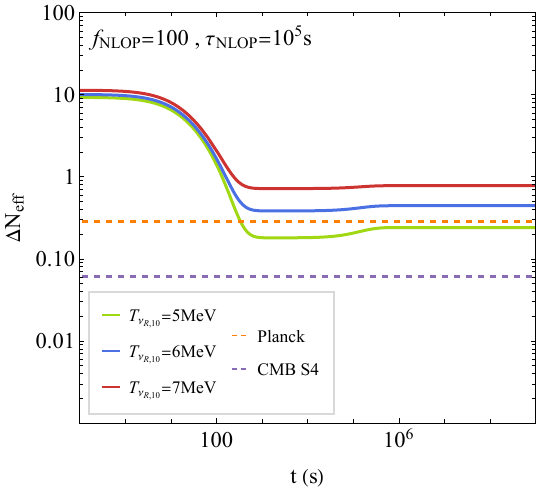}
    \includegraphics[width=0.45\linewidth]{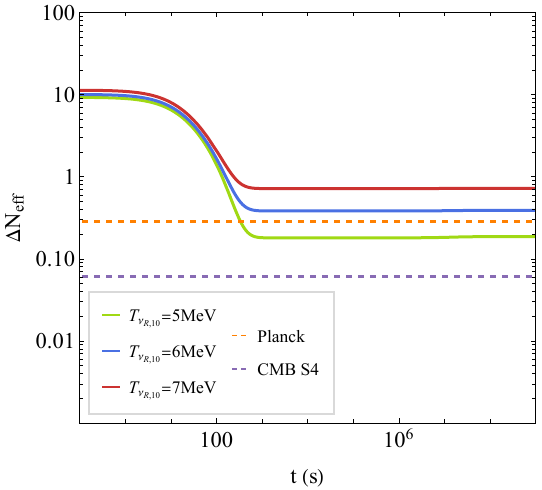}
    \caption{Evolution of $\Delta N_{\mathrm{eff}}$ for varying $\tau_{\mathrm{NLOP}}$, $f_{\mathrm{NLOP}}$, and $T_{\nu_R,10}$. The final subfigure illustrates the scenario in which $\Delta N_{\mathrm{eff}}$ is dominated by thermal contributions.}
    \label{fig:Neff}
\end{figure}

With the time-dependent temperatures $T_{\nu_L}$, $T_{\nu_R}$, and $T_\gamma$ determined, $\Delta N_{\mathrm{eff}}$ can be computed using Eq.~\eqref{eq:Neff}. 
From Eq.~\eqref{eq:Tode}, it follows that the energy injection from NLOP decay into $\nu_L$ or $\nu_R$ proceeds in an identical manner in both scenarios. 
As a result, the evolution of $\Delta N_{\mathrm{eff}}$ is the same whether the NLOP is $\phi_R$ or $\chi$. 
We present the resulting $\Delta N_{\mathrm{eff}}$ in Fig.~\ref{fig:Neff}, where the three subfigures show the effects of varying $\tau_{\mathrm{NLOP}}$, $f_{\mathrm{NLOP}}$, and $T_{\nu_R,10}$, respectively. The two dashed lines stand for the current experimental limits from Planck~\cite{Planck:2018vyg} and future detecting capability from CMB-S4~\cite{CMB-S4:2016ple}.
A longer NLOP lifetime or a larger $f_{\mathrm{NLOP}}$ leads to a more pronounced heating effect, resulting in a higher $\Delta N_{\mathrm{eff}}$. 
A hotter $\nu_R$ bath, originating from stronger and more prolonged interactions between the dark plasma and the SM plasma, also yields a larger $\Delta N_{\mathrm{eff}}$. 
For completeness, we also show the evolution of $\Delta N_{\mathrm{eff}}$ in the scenario where the contribution from the late decay of the NLOP is negligible, as depicted in the final subfigure of Fig.~\ref{fig:Neff}. This occurs when either the NLOP abundance is small or the NLOP lifetime is sufficiently short. In this limit, $\Delta N_{\mathrm{eff}}$ is dominated solely by the thermal contribution from the decoupled $\nu_R$ bath.
As expected, a higher $T_{\nu_R,10}$ results in a larger $\Delta N_{\mathrm{eff}}$, since the $\nu_R$s contribute more to the radiation density.

\subsection{Combined Results}
Finally, we perform a comprehensive scan over the full parameter space, imposing constraints from dark matter relic density, neutrino mass, LFV, and cosmological observations such as BBN, CMB, and $\Delta N_{\mathrm{eff}}$. The scanning ranges for the model parameters are as follows:
\begin{align}
m_{N_1} &\in [1, 1000] \, \text{GeV}, 
& m_{\text{NLOP}} &\in [m_{N_1}, 1000] \, \text{GeV}, 
& m_{\text{others}} &\in [m_{\text{NLOP}}, 2000] \, \text{GeV}, \notag \\
(y_{\Phi,\chi})_{\alpha 1} &\in [10^{-15}, 10^{-7}], 
& (y_{\Phi})_{\alpha 2,3} &\in [10^{-5}, 10^{-2}], 
& (y_{\chi})_{\alpha 2,3} &\in [10^{-3}, 1], \\
\lambda_5 &\in [10^{-3}, 1], 
& \sin 2\theta &\in [10^{-10}, 10^{-5}]. \notag
\end{align}
Here $m_{\text{NLOP}}$ is the mass of the NLOP, and $m_{\text{others}}$ refers to the masses of heavier odd-sector states. The lower bound on $m_{\phi^{\pm}}$ is set to \SI{560}{GeV}, based on collider searches for long-lived charged particles~\cite{ATLAS:2025fdm}. The dark matter relic density is required to lie within the $3\sigma$ range of the Planck2018 measurement~\cite{Planck:2018vyg}, i.e., $\Omega_{N_1} h^2 \in [0.117, 0.123]$. The odd states masses are scanned up to energy scale within reach of current and near-future experiments. The couplings $(y_{\Phi,\chi})_{\alpha 1}$ are bounded from above to ensure that dark matter remains out of thermal equilibrium. The couplings $(y_{\Phi})_{\alpha 2,3}$ are constrained by LFV bounds to be less than $10^{-2}$, while $(y_{\chi})_{\alpha 2,3}$ are chosen to reproduce observed neutrino masses. The mixing parameter $\sin 2\theta$ is restricted to below $10^{-5}$ on naturalness grounds—since $\theta = 0$ (or equivalently $\mu = 0$) enhances the model's symmetry. The quartic coupling $\lambda_5$ is scanned down to $10^{-3}$, as the thermal contribution to $\Delta N_{\rm eff}$ is no longer significant for smaller values.

\begin{figure}
    \centering
    \includegraphics[width=0.5\linewidth]{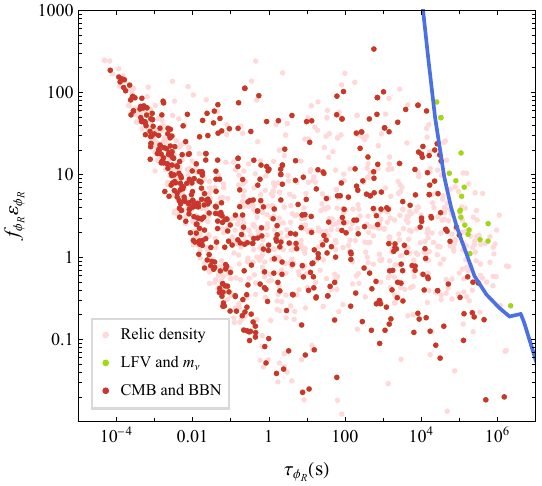}
    \caption{Scan results in the $f_{\phi_R} \varepsilon_{\phi_R}$--$\tau_{\phi_R}$ plane for the case where $\phi_R$ is the NLOP. The blue curve shows the combined constraints from BBN and CMB as derived in Ref.~\cite{Hambye:2021moy}.}
    \label{fig:CMBBBN}
\end{figure}

As discussed in the previous subsection, we impose a conservative upper limit of $\tau < \SI{3e4}{\second}$ on the lifetime of the NLOP when it is either the charged scalar $\phi^\pm$ or the neutral fermion $N_2$. This bound arises because EM energy injection from late decays can alter light element abundances during BBN and distort CMB anisotropies. However, when the NLOP is the neutral scalar $\phi_R$, the constraints are significantly relaxed, as its dominant decays inject energy primarily into neutrinos rather than the EM plasma. In this case, we adopt the bounds from Ref.~\cite{Hambye:2021moy}, which depend on the effective energy transfer parameter $f_{\phi_R} \varepsilon_{\phi_R}$, and display them as the blue curve in Fig.~\ref{fig:CMBBBN}.
In the scan, parameter points satisfying the dark matter relic density condition are colored pink. Among these, points that additionally satisfy constraints from neutrino mass and LFV processes are highlighted in green. Points that further pass the BBN and CMB bounds are shown in red. One can observe that the BBN and CMB limits exclude long-lived $\phi_R$ with lifetimes $\gtrsim 10^4$--$10^7\,\si{\second}$, although the constraints weaken significantly when the energy transferred to the neutrino sector is small.
When the NLOP is the scalar $\chi$, which decays directly into right-handed neutrino, there are no direct BBN or CMB constraints, as the decay products are sterile and do not interact with the thermal bath.

\begin{figure}
    \centering
    \includegraphics[width=0.48\linewidth]{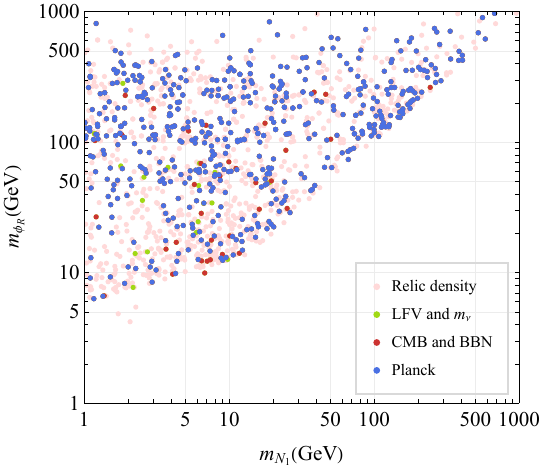}
    \includegraphics[width=0.48\linewidth]{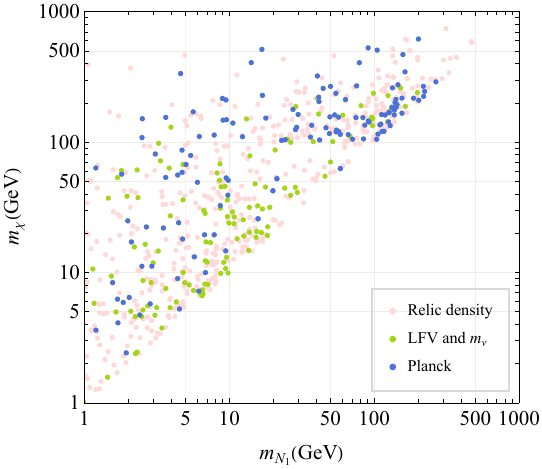}
    \includegraphics[width=0.48\linewidth]{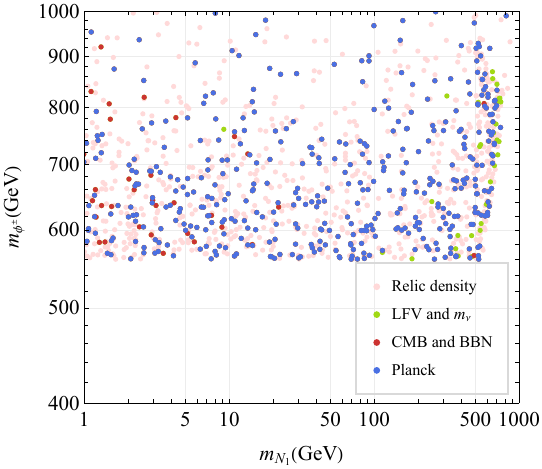} 
    \includegraphics[width=0.48\linewidth]{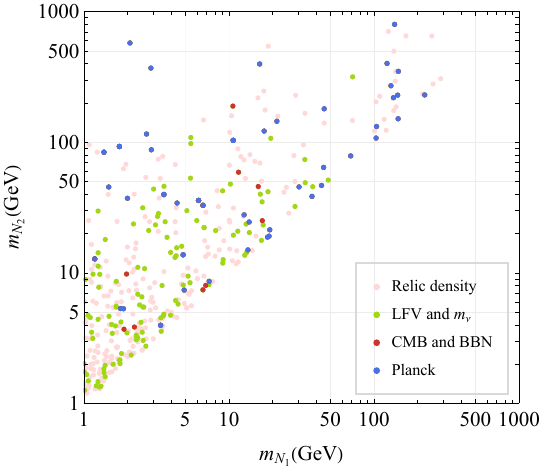}
    \caption{Results of parameter scan in the $m_{\rm NLOP}-m_{N_1}$ plane under combined constraints from DM relic density, LFV, neutrino mass, CMB, BBN, and $\Delta N_{\rm eff}$.}
    \label{fig:combine}
\end{figure}

In Fig.~\ref{fig:combine}, we present the scanning results in the $m_{\mathrm{NLOP}}$-$m_{N_1}$ plane, considering different particles as the NLOP. Constraints from dark matter relic density, neutrino mass and LFV, BBN and CMB, and $\Delta N_{\mathrm{eff}}$ are applied successively. The color coding follows the same progression as in Fig.~\ref{fig:CMBBBN}. Finally, points that also satisfy the $\Delta N_{\mathrm{eff}}$ constraint are marked in blue, indicating full consistency with all observational requirements.
The dark matter relic density receives contributions from both the freeze-in and super-WIMP mechanisms, as discussed in detail in Sect.~\ref{sec:relic}. Constraints from LFV, particularly the stringent experimental limit on the branching ratio of $\mu \to e\gamma$, exclude a significant portion of the parameter space. The impact of LFV constraints is also visible in the subfigure where $\chi$ is the NLOP, even though $\chi$ itself does not directly participate in LFV processes. This occurs because the same Yukawa couplings that govern LFV amplitudes also influence the dark matter production rate and thus must satisfy both relic density and flavor constraints simultaneously.
The bounds from BBN and CMB have been discussed previously and are applied depending on the identity of the NLOP.
In all cases—whether $\phi_R$, $\chi$, $\phi^\pm$, or $N_2$ serves as the NLOP—we identify regions of parameter space that satisfy all observational constraints, including the latest limits on $\Delta N_{\mathrm{eff}}$. This demonstrates the viability of the FIMP Dirac scotogenic Model as a unified framework for explaining both neutrino mass generation and the observed dark matter relic abundance.
Future experiments will play a crucial role in testing this scenario. Long-lived particle searches at proposed facilities such as MATHUSLA~\cite{Curtin:2018mvb} could probe the metastable NLOP states, while next-generation CMB observations from CMB-S4~\cite{CMB-S4:2016ple} will improve sensitivity to $\Delta N_{\mathrm{eff}}$ and energy injection during cosmic evolution. These experiments will either confirm the predictions of the model or further constrain its surviving parameter space.

\section{Conclusion}\label{sec:conclusion}

In this work, we have studied the FIMP Dirac Scotogenic Model, a well-motivated extension of the SM that simultaneously addresses the origin of neutrino masses and the nature of dark matter. The lightest neutral fermion $N_1$ is a feebly interacting massive particle DM candidate, produced out of equilibrium. Neutrino masses are generated radiatively at one loop, resulting in a rank-2 mass matrix due to highly suppressed couplings to $N_1$, predicting one nearly massless neutrino. Stringent LFV constraints, require the Yukawa couplings $(y_\Phi)_{\alpha 2,3}$ to be small, which in turn necessitates larger $(y_\chi)_{\alpha 2,3}$ to fit the neutrino mass scale. This coupling structure crucially impacts the phenomenology, governing the freeze-out dynamics of the NLOP and the thermal history of the right-handed neutrino bath.

We have systematically analyzed two distinct production mechanisms for $N_1$: direct freeze-in via decays of heavier $Z_2$-odd scalars and fermions, and the ``super-WIMP'' mechanism, where $N_1$ is produced from the late decay of a thermally frozen-out NLOP. The latter scenario leads to rich phenomenology, depending on whether $N_2$, $\phi_{R,I}$, $\phi^\pm$, or $\chi$ serves as the NLOP. We demonstrate that the correct relic abundance can be achieved across various benchmarks, with coannihilation effects and enhanced annihilation channels playing crucial roles in regulating the NLOP's freeze-out density, offering flexible pathways to match the observed DM density.

Another central focus of this work is the cosmological impact on the effective number of relativistic species. Our analysis reveals two distinct contributions to this quantity. The first is a thermal contribution, arising from the primordial right-handed neutrino bath. This component is determined by the decoupling temperature of the $\nu_R$ bath, which is set by the strength of interactions between the dark sector and the SM plasma. The second contribution is non-thermal, originating from the late decays of the NLOP. The magnitude of this non-thermal contribution is highly sensitive to the NLOP's properties: both a longer lifetime $\tau_{\mathrm{NLOP}}$ and a larger energy fraction $f_{\mathrm{NLOP}}$ lead to a more pronounced heating effect and thus a higher $\Delta N_{\mathrm{eff}}$.

We have performed a comprehensive parameter scan, sequentially applying constraints from DM relic density, neutrino mass and LFV, BBN/CMB, and $\Delta N_{\mathrm{eff}}$. Viable parameter space exists for all NLOP candidates that satisfy all current observational bounds. Future experiments setup for long-lived particle searches and precision $\Delta N_{\mathrm{eff}}$ measurements will provide powerful probes to either discover this framework or further constrain its remaining parameter space.

Finally, It is instructive to contrast the two possible dark matter production mechanisms in our work: thermal freeze-out (WIMP)~\cite{Guo:2020qin} versus freeze-in (FIMP). The distinction is most pronounced in the size of the Yukawa couplings $(y_{\Phi,\chi})_{\alpha 1}$ that govern the interaction of the lightest odd particle $N_1$ with the Standard Model sector. In the WIMP scenario, these couplings must be relatively large to ensure efficient annihilation and avoid overproduction of dark matter. In the FIMP regime, by contrast, the couplings are extremely suppressed, $(y_{\Phi,\chi})_{\alpha 1} \lesssim 10^{-7}$, keeping $N_1$ out of thermal equilibrium throughout cosmic history. The relic density is then generated via the slow decay of heavier states, and this suppression naturally leads to a rank-2 neutrino mass matrix, predicting one nearly massless neutrino—a feature not generic in WIMP realizations.
Phenomenologically, the two scenarios are also sharply distinguished by experimental constraints. WIMP parameter space is heavily constrained: collider searches exclude light dark matter masses, while direct and indirect detection experiments probe the TeV-scale region. In contrast, FIMPs evade these conventional probes due to their feeble couplings. Their collider signatures are limited to displaced decays of the NLOP (e.g., displaced vertices), and current long-lived particle searches provide only weak limits. Direct and indirect DM detection signals are negligible. Instead, the primary observational window for FIMPs lies in cosmology—particularly through their contribution to $\Delta N_{\mathrm{eff}}$ from late decays of long-lived particles—which offers a powerful and complementary test of the freeze-in paradigm.

\section*{Acknowledgments}
We thank Zhi-Long Han, Ang Liu and Xiao-Dong Ma for helpful discussions. This work by S.-Y.G. was supported by the NNSFC under Grant No. 12305113, by the Natural Science Foundation of Shandong Province under Grants No. ZR2022QA026, and by the Project of Shandong Province Higher Educational Science and Technology
Program under Grants No. 2022KJ271.

\bibliography{main}

\begin{thebibliography}{85}
\expandafter\ifx\csname natexlab\endcsname\relax\def\natexlab#1{#1}\fi
\expandafter\ifx\csname bibnamefont\endcsname\relax
  \def\bibnamefont#1{#1}\fi
\expandafter\ifx\csname bibfnamefont\endcsname\relax
  \def\bibfnamefont#1{#1}\fi
\expandafter\ifx\csname citenamefont\endcsname\relax
  \def\citenamefont#1{#1}\fi
\expandafter\ifx\csname url\endcsname\relax
  \def\url#1{\texttt{#1}}\fi
\expandafter\ifx\csname urlprefix\endcsname\relax\def\urlprefix{URL }\fi
\providecommand{\bibinfo}[2]{#2}
\providecommand{\eprint}[2][]{\url{#2}}

\bibitem[{\citenamefont{Davis et~al.}(1968)\citenamefont{Davis, Harmer, and Hoffman}}]{Davis:1968cp}
\bibinfo{author}{\bibfnamefont{R.}~\bibnamefont{Davis}, \bibfnamefont{Jr.}}, \bibinfo{author}{\bibfnamefont{D.~S.} \bibnamefont{Harmer}}, \bibnamefont{and} \bibinfo{author}{\bibfnamefont{K.~C.} \bibnamefont{Hoffman}}, \bibinfo{journal}{Phys. Rev. Lett.} \textbf{\bibinfo{volume}{20}}, \bibinfo{pages}{1205} (\bibinfo{year}{1968}).

\bibitem[{\citenamefont{Fukuda et~al.}(1998)}]{Super-Kamiokande:1998kpq}
\bibinfo{author}{\bibfnamefont{Y.}~\bibnamefont{Fukuda}} \bibnamefont{et~al.} (\bibinfo{collaboration}{Super-Kamiokande}), \bibinfo{journal}{Phys. Rev. Lett.} \textbf{\bibinfo{volume}{81}}, \bibinfo{pages}{1562} (\bibinfo{year}{1998}), \eprint{hep-ex/9807003}.

\bibitem[{\citenamefont{An et~al.}(2012)}]{DayaBay:2012fng}
\bibinfo{author}{\bibfnamefont{F.~P.} \bibnamefont{An}} \bibnamefont{et~al.} (\bibinfo{collaboration}{Daya Bay}), \bibinfo{journal}{Phys. Rev. Lett.} \textbf{\bibinfo{volume}{108}}, \bibinfo{pages}{171803} (\bibinfo{year}{2012}), \eprint{1203.1669}.

\bibitem[{\citenamefont{Ahn et~al.}(2012)}]{RENO:2012mkc}
\bibinfo{author}{\bibfnamefont{J.~K.} \bibnamefont{Ahn}} \bibnamefont{et~al.} (\bibinfo{collaboration}{RENO}), \bibinfo{journal}{Phys. Rev. Lett.} \textbf{\bibinfo{volume}{108}}, \bibinfo{pages}{191802} (\bibinfo{year}{2012}), \eprint{1204.0626}.

\bibitem[{\citenamefont{Aliu et~al.}(2005)}]{K2K:2004iot}
\bibinfo{author}{\bibfnamefont{E.}~\bibnamefont{Aliu}} \bibnamefont{et~al.} (\bibinfo{collaboration}{K2K}), \bibinfo{journal}{Phys. Rev. Lett.} \textbf{\bibinfo{volume}{94}}, \bibinfo{pages}{081802} (\bibinfo{year}{2005}), \eprint{hep-ex/0411038}.

\bibitem[{\citenamefont{Abe et~al.}(2012)}]{DoubleChooz:2011ymz}
\bibinfo{author}{\bibfnamefont{Y.}~\bibnamefont{Abe}} \bibnamefont{et~al.} (\bibinfo{collaboration}{Double Chooz}), \bibinfo{journal}{Phys. Rev. Lett.} \textbf{\bibinfo{volume}{108}}, \bibinfo{pages}{131801} (\bibinfo{year}{2012}), \eprint{1112.6353}.

\bibitem[{\citenamefont{Cai et~al.}(2018)\citenamefont{Cai, Han, Li, and Ruiz}}]{Cai:2017mow}
\bibinfo{author}{\bibfnamefont{Y.}~\bibnamefont{Cai}}, \bibinfo{author}{\bibfnamefont{T.}~\bibnamefont{Han}}, \bibinfo{author}{\bibfnamefont{T.}~\bibnamefont{Li}}, \bibnamefont{and} \bibinfo{author}{\bibfnamefont{R.}~\bibnamefont{Ruiz}}, \bibinfo{journal}{Front. in Phys.} \textbf{\bibinfo{volume}{6}}, \bibinfo{pages}{40} (\bibinfo{year}{2018}), \eprint{1711.02180}.

\bibitem[{\citenamefont{Sirunyan et~al.}(2019)}]{CMS:2018jxx}
\bibinfo{author}{\bibfnamefont{A.~M.} \bibnamefont{Sirunyan}} \bibnamefont{et~al.} (\bibinfo{collaboration}{CMS}), \bibinfo{journal}{JHEP} \textbf{\bibinfo{volume}{01}}, \bibinfo{pages}{122} (\bibinfo{year}{2019}), \eprint{1806.10905}.

\bibitem[{\citenamefont{Dolinski et~al.}(2019)\citenamefont{Dolinski, Poon, and Rodejohann}}]{Dolinski:2019nrj}
\bibinfo{author}{\bibfnamefont{M.~J.} \bibnamefont{Dolinski}}, \bibinfo{author}{\bibfnamefont{A.~W.~P.} \bibnamefont{Poon}}, \bibnamefont{and} \bibinfo{author}{\bibfnamefont{W.}~\bibnamefont{Rodejohann}}, \bibinfo{journal}{Ann. Rev. Nucl. Part. Sci.} \textbf{\bibinfo{volume}{69}}, \bibinfo{pages}{219} (\bibinfo{year}{2019}), \eprint{1902.04097}.

\bibitem[{\citenamefont{Anton et~al.}(2019)}]{EXO-200:2019rkq}
\bibinfo{author}{\bibfnamefont{G.}~\bibnamefont{Anton}} \bibnamefont{et~al.} (\bibinfo{collaboration}{EXO-200}), \bibinfo{journal}{Phys. Rev. Lett.} \textbf{\bibinfo{volume}{123}}, \bibinfo{pages}{161802} (\bibinfo{year}{2019}), \eprint{1906.02723}.

\bibitem[{\citenamefont{Abgrall et~al.}(2021)}]{LEGEND:2021bnm}
\bibinfo{author}{\bibfnamefont{N.}~\bibnamefont{Abgrall}} \bibnamefont{et~al.} (\bibinfo{collaboration}{LEGEND}) (\bibinfo{year}{2021}), \eprint{2107.11462}.

\bibitem[{\citenamefont{Alvis et~al.}(2019)}]{Majorana:2019nbd}
\bibinfo{author}{\bibfnamefont{S.~I.} \bibnamefont{Alvis}} \bibnamefont{et~al.} (\bibinfo{collaboration}{Majorana}), \bibinfo{journal}{Phys. Rev. C} \textbf{\bibinfo{volume}{100}}, \bibinfo{pages}{025501} (\bibinfo{year}{2019}), \eprint{1902.02299}.

\bibitem[{\citenamefont{Armengaud et~al.}(2021)}]{CUPID:2020aow}
\bibinfo{author}{\bibfnamefont{E.}~\bibnamefont{Armengaud}} \bibnamefont{et~al.} (\bibinfo{collaboration}{CUPID}), \bibinfo{journal}{Phys. Rev. Lett.} \textbf{\bibinfo{volume}{126}}, \bibinfo{pages}{181802} (\bibinfo{year}{2021}), \eprint{2011.13243}.

\bibitem[{\citenamefont{Abe et~al.}(2023)}]{KamLAND-Zen:2022tow}
\bibinfo{author}{\bibfnamefont{S.}~\bibnamefont{Abe}} \bibnamefont{et~al.} (\bibinfo{collaboration}{KamLAND-Zen}), \bibinfo{journal}{Phys. Rev. Lett.} \textbf{\bibinfo{volume}{130}}, \bibinfo{pages}{051801} (\bibinfo{year}{2023}), \eprint{2203.02139}.

\bibitem[{\citenamefont{Yao and Ding}(2018)}]{Yao:2018ekp}
\bibinfo{author}{\bibfnamefont{C.-Y.} \bibnamefont{Yao}} \bibnamefont{and} \bibinfo{author}{\bibfnamefont{G.-J.} \bibnamefont{Ding}}, \bibinfo{journal}{Phys. Rev. D} \textbf{\bibinfo{volume}{97}}, \bibinfo{pages}{095042} (\bibinfo{year}{2018}), \eprint{1802.05231}.

\bibitem[{\citenamefont{Centelles~Chuli{\'a} et~al.}(2018{\natexlab{a}})\citenamefont{Centelles~Chuli{\'a}, Srivastava, and Valle}}]{CentellesChulia:2018bkz}
\bibinfo{author}{\bibfnamefont{S.}~\bibnamefont{Centelles~Chuli{\'a}}}, \bibinfo{author}{\bibfnamefont{R.}~\bibnamefont{Srivastava}}, \bibnamefont{and} \bibinfo{author}{\bibfnamefont{J.~W.~F.} \bibnamefont{Valle}}, \bibinfo{journal}{Phys. Rev. D} \textbf{\bibinfo{volume}{98}}, \bibinfo{pages}{035009} (\bibinfo{year}{2018}{\natexlab{a}}), \eprint{1804.03181}.

\bibitem[{\citenamefont{Calle et~al.}(2019)\citenamefont{Calle, Restrepo, Yaguna, and Zapata}}]{Calle:2018ovc}
\bibinfo{author}{\bibfnamefont{J.}~\bibnamefont{Calle}}, \bibinfo{author}{\bibfnamefont{D.}~\bibnamefont{Restrepo}}, \bibinfo{author}{\bibfnamefont{C.~E.} \bibnamefont{Yaguna}}, \bibnamefont{and} \bibinfo{author}{\bibfnamefont{{\'O}.}~\bibnamefont{Zapata}}, \bibinfo{journal}{Phys. Rev. D} \textbf{\bibinfo{volume}{99}}, \bibinfo{pages}{075008} (\bibinfo{year}{2019}), \eprint{1812.05523}.

\bibitem[{\citenamefont{Jana et~al.}(2020)\citenamefont{Jana, Vishnu, and Saad}}]{Jana:2019mgj}
\bibinfo{author}{\bibfnamefont{S.}~\bibnamefont{Jana}}, \bibinfo{author}{\bibfnamefont{P.~K.} \bibnamefont{Vishnu}}, \bibnamefont{and} \bibinfo{author}{\bibfnamefont{S.}~\bibnamefont{Saad}}, \bibinfo{journal}{JCAP} \textbf{\bibinfo{volume}{04}}, \bibinfo{pages}{018} (\bibinfo{year}{2020}), \eprint{1910.09537}.

\bibitem[{\citenamefont{Saad}(2019)}]{Saad:2019bqf}
\bibinfo{author}{\bibfnamefont{S.}~\bibnamefont{Saad}}, \bibinfo{journal}{Nucl. Phys. B} \textbf{\bibinfo{volume}{943}}, \bibinfo{pages}{114636} (\bibinfo{year}{2019}), \eprint{1902.07259}.

\bibitem[{\citenamefont{Farzan and Ma}(2012)}]{Farzan:2012sa}
\bibinfo{author}{\bibfnamefont{Y.}~\bibnamefont{Farzan}} \bibnamefont{and} \bibinfo{author}{\bibfnamefont{E.}~\bibnamefont{Ma}}, \bibinfo{journal}{Phys. Rev. D} \textbf{\bibinfo{volume}{86}}, \bibinfo{pages}{033007} (\bibinfo{year}{2012}), \eprint{1204.4890}.

\bibitem[{\citenamefont{Gu and He}(2006)}]{Gu:2006dc}
\bibinfo{author}{\bibfnamefont{P.-H.} \bibnamefont{Gu}} \bibnamefont{and} \bibinfo{author}{\bibfnamefont{H.-J.} \bibnamefont{He}}, \bibinfo{journal}{JCAP} \textbf{\bibinfo{volume}{12}}, \bibinfo{pages}{010} (\bibinfo{year}{2006}), \eprint{hep-ph/0610275}.

\bibitem[{\citenamefont{Gu and Sarkar}(2008)}]{Gu:2007ug}
\bibinfo{author}{\bibfnamefont{P.-H.} \bibnamefont{Gu}} \bibnamefont{and} \bibinfo{author}{\bibfnamefont{U.}~\bibnamefont{Sarkar}}, \bibinfo{journal}{Phys. Rev. D} \textbf{\bibinfo{volume}{77}}, \bibinfo{pages}{105031} (\bibinfo{year}{2008}), \eprint{0712.2933}.

\bibitem[{\citenamefont{Centelles~Chuli{\'a} et~al.}(2017{\natexlab{a}})\citenamefont{Centelles~Chuli{\'a}, Ma, Srivastava, and Valle}}]{Chulia:2016ngi}
\bibinfo{author}{\bibfnamefont{S.}~\bibnamefont{Centelles~Chuli{\'a}}}, \bibinfo{author}{\bibfnamefont{E.}~\bibnamefont{Ma}}, \bibinfo{author}{\bibfnamefont{R.}~\bibnamefont{Srivastava}}, \bibnamefont{and} \bibinfo{author}{\bibfnamefont{J.~W.~F.} \bibnamefont{Valle}}, \bibinfo{journal}{Phys. Lett. B} \textbf{\bibinfo{volume}{767}}, \bibinfo{pages}{209} (\bibinfo{year}{2017}{\natexlab{a}}), \eprint{1606.04543}.

\bibitem[{\citenamefont{Bonilla et~al.}(2016)\citenamefont{Bonilla, Ma, Peinado, and Valle}}]{Bonilla:2016diq}
\bibinfo{author}{\bibfnamefont{C.}~\bibnamefont{Bonilla}}, \bibinfo{author}{\bibfnamefont{E.}~\bibnamefont{Ma}}, \bibinfo{author}{\bibfnamefont{E.}~\bibnamefont{Peinado}}, \bibnamefont{and} \bibinfo{author}{\bibfnamefont{J.~W.~F.} \bibnamefont{Valle}}, \bibinfo{journal}{Phys. Lett. B} \textbf{\bibinfo{volume}{762}}, \bibinfo{pages}{214} (\bibinfo{year}{2016}), \eprint{1607.03931}.

\bibitem[{\citenamefont{Wang and Han}(2017)}]{Wang:2016lve}
\bibinfo{author}{\bibfnamefont{W.}~\bibnamefont{Wang}} \bibnamefont{and} \bibinfo{author}{\bibfnamefont{Z.-L.} \bibnamefont{Han}}, \bibinfo{journal}{JHEP} \textbf{\bibinfo{volume}{04}}, \bibinfo{pages}{166} (\bibinfo{year}{2017}), \eprint{1611.03240}.

\bibitem[{\citenamefont{Borah and Dasgupta}(2017)}]{Borah:2017leo}
\bibinfo{author}{\bibfnamefont{D.}~\bibnamefont{Borah}} \bibnamefont{and} \bibinfo{author}{\bibfnamefont{A.}~\bibnamefont{Dasgupta}}, \bibinfo{journal}{JCAP} \textbf{\bibinfo{volume}{06}}, \bibinfo{pages}{003} (\bibinfo{year}{2017}), \eprint{1702.02877}.

\bibitem[{\citenamefont{Wang et~al.}(2017)\citenamefont{Wang, Wang, Han, and Han}}]{Wang:2017mcy}
\bibinfo{author}{\bibfnamefont{W.}~\bibnamefont{Wang}}, \bibinfo{author}{\bibfnamefont{R.}~\bibnamefont{Wang}}, \bibinfo{author}{\bibfnamefont{Z.-L.} \bibnamefont{Han}}, \bibnamefont{and} \bibinfo{author}{\bibfnamefont{J.-Z.} \bibnamefont{Han}}, \bibinfo{journal}{Eur. Phys. J. C} \textbf{\bibinfo{volume}{77}}, \bibinfo{pages}{889} (\bibinfo{year}{2017}), \eprint{1705.00414}.

\bibitem[{\citenamefont{Centelles~Chuli{\'a} et~al.}(2017{\natexlab{b}})\citenamefont{Centelles~Chuli{\'a}, Srivastava, and Valle}}]{CentellesChulia:2017koy}
\bibinfo{author}{\bibfnamefont{S.}~\bibnamefont{Centelles~Chuli{\'a}}}, \bibinfo{author}{\bibfnamefont{R.}~\bibnamefont{Srivastava}}, \bibnamefont{and} \bibinfo{author}{\bibfnamefont{J.~W.~F.} \bibnamefont{Valle}}, \bibinfo{journal}{Phys. Lett. B} \textbf{\bibinfo{volume}{773}}, \bibinfo{pages}{26} (\bibinfo{year}{2017}{\natexlab{b}}), \eprint{1706.00210}.

\bibitem[{\citenamefont{Ma and Sarkar}(2018)}]{Ma:2017kgb}
\bibinfo{author}{\bibfnamefont{E.}~\bibnamefont{Ma}} \bibnamefont{and} \bibinfo{author}{\bibfnamefont{U.}~\bibnamefont{Sarkar}}, \bibinfo{journal}{Phys. Lett. B} \textbf{\bibinfo{volume}{776}}, \bibinfo{pages}{54} (\bibinfo{year}{2018}), \eprint{1707.07698}.

\bibitem[{\citenamefont{Yao and Ding}(2017)}]{Yao:2017vtm}
\bibinfo{author}{\bibfnamefont{C.-Y.} \bibnamefont{Yao}} \bibnamefont{and} \bibinfo{author}{\bibfnamefont{G.-J.} \bibnamefont{Ding}}, \bibinfo{journal}{Phys. Rev. D} \textbf{\bibinfo{volume}{96}}, \bibinfo{pages}{095004} (\bibinfo{year}{2017}), \bibinfo{note}{[Erratum: Phys.Rev.D 98, 039901 (2018)]}, \eprint{1707.09786}.

\bibitem[{\citenamefont{Bonilla et~al.}(2018)\citenamefont{Bonilla, Lamprea, Peinado, and Valle}}]{Bonilla:2017ekt}
\bibinfo{author}{\bibfnamefont{C.}~\bibnamefont{Bonilla}}, \bibinfo{author}{\bibfnamefont{J.~M.} \bibnamefont{Lamprea}}, \bibinfo{author}{\bibfnamefont{E.}~\bibnamefont{Peinado}}, \bibnamefont{and} \bibinfo{author}{\bibfnamefont{J.~W.~F.} \bibnamefont{Valle}}, \bibinfo{journal}{Phys. Lett. B} \textbf{\bibinfo{volume}{779}}, \bibinfo{pages}{257} (\bibinfo{year}{2018}), \eprint{1710.06498}.

\bibitem[{\citenamefont{Ibarra et~al.}(2018)\citenamefont{Ibarra, Kushwaha, and Vempati}}]{Ibarra:2017tju}
\bibinfo{author}{\bibfnamefont{A.}~\bibnamefont{Ibarra}}, \bibinfo{author}{\bibfnamefont{A.}~\bibnamefont{Kushwaha}}, \bibnamefont{and} \bibinfo{author}{\bibfnamefont{S.~K.} \bibnamefont{Vempati}}, \bibinfo{journal}{Phys. Lett. B} \textbf{\bibinfo{volume}{780}}, \bibinfo{pages}{86} (\bibinfo{year}{2018}), \eprint{1711.02070}.

\bibitem[{\citenamefont{Borah and Karmakar}(2018)}]{Borah:2017dmk}
\bibinfo{author}{\bibfnamefont{D.}~\bibnamefont{Borah}} \bibnamefont{and} \bibinfo{author}{\bibfnamefont{B.}~\bibnamefont{Karmakar}}, \bibinfo{journal}{Phys. Lett. B} \textbf{\bibinfo{volume}{780}}, \bibinfo{pages}{461} (\bibinfo{year}{2018}), \eprint{1712.06407}.

\bibitem[{\citenamefont{Das et~al.}(2017)\citenamefont{Das, Nomura, Okada, and Roy}}]{Das:2017ski}
\bibinfo{author}{\bibfnamefont{A.}~\bibnamefont{Das}}, \bibinfo{author}{\bibfnamefont{T.}~\bibnamefont{Nomura}}, \bibinfo{author}{\bibfnamefont{H.}~\bibnamefont{Okada}}, \bibnamefont{and} \bibinfo{author}{\bibfnamefont{S.}~\bibnamefont{Roy}}, \bibinfo{journal}{Phys. Rev. D} \textbf{\bibinfo{volume}{96}}, \bibinfo{pages}{075001} (\bibinfo{year}{2017}), \eprint{1704.02078}.

\bibitem[{\citenamefont{Centelles~Chuli{\'a} et~al.}(2018{\natexlab{b}})\citenamefont{Centelles~Chuli{\'a}, Srivastava, and Valle}}]{CentellesChulia:2018gwr}
\bibinfo{author}{\bibfnamefont{S.}~\bibnamefont{Centelles~Chuli{\'a}}}, \bibinfo{author}{\bibfnamefont{R.}~\bibnamefont{Srivastava}}, \bibnamefont{and} \bibinfo{author}{\bibfnamefont{J.~W.~F.} \bibnamefont{Valle}}, \bibinfo{journal}{Phys. Lett. B} \textbf{\bibinfo{volume}{781}}, \bibinfo{pages}{122} (\bibinfo{year}{2018}{\natexlab{b}}), \eprint{1802.05722}.

\bibitem[{\citenamefont{Han and Wang}(2018)}]{Han:2018zcn}
\bibinfo{author}{\bibfnamefont{Z.-L.} \bibnamefont{Han}} \bibnamefont{and} \bibinfo{author}{\bibfnamefont{W.}~\bibnamefont{Wang}}, \bibinfo{journal}{Eur. Phys. J. C} \textbf{\bibinfo{volume}{78}}, \bibinfo{pages}{839} (\bibinfo{year}{2018}), \eprint{1805.02025}.

\bibitem[{\citenamefont{Borah et~al.}(2018)\citenamefont{Borah, Karmakar, and Nanda}}]{Borah:2018gjk}
\bibinfo{author}{\bibfnamefont{D.}~\bibnamefont{Borah}}, \bibinfo{author}{\bibfnamefont{B.}~\bibnamefont{Karmakar}}, \bibnamefont{and} \bibinfo{author}{\bibfnamefont{D.}~\bibnamefont{Nanda}}, \bibinfo{journal}{JCAP} \textbf{\bibinfo{volume}{07}}, \bibinfo{pages}{039} (\bibinfo{year}{2018}), \eprint{1805.11115}.

\bibitem[{\citenamefont{Borah and Karmakar}(2019)}]{Borah:2018nvu}
\bibinfo{author}{\bibfnamefont{D.}~\bibnamefont{Borah}} \bibnamefont{and} \bibinfo{author}{\bibfnamefont{B.}~\bibnamefont{Karmakar}}, \bibinfo{journal}{Phys. Lett. B} \textbf{\bibinfo{volume}{789}}, \bibinfo{pages}{59} (\bibinfo{year}{2019}), \eprint{1806.10685}.

\bibitem[{\citenamefont{Carvajal and Zapata}(2019)}]{Carvajal:2018ohk}
\bibinfo{author}{\bibfnamefont{C.~D.~R.} \bibnamefont{Carvajal}} \bibnamefont{and} \bibinfo{author}{\bibfnamefont{{\'O}.}~\bibnamefont{Zapata}}, \bibinfo{journal}{Phys. Rev. D} \textbf{\bibinfo{volume}{99}}, \bibinfo{pages}{075009} (\bibinfo{year}{2019}), \eprint{1812.06364}.

\bibitem[{\citenamefont{Ma}(2019{\natexlab{a}})}]{Ma:2019yfo}
\bibinfo{author}{\bibfnamefont{E.}~\bibnamefont{Ma}}, \bibinfo{journal}{Phys. Lett. B} \textbf{\bibinfo{volume}{793}}, \bibinfo{pages}{411} (\bibinfo{year}{2019}{\natexlab{a}}), \eprint{1901.09091}.

\bibitem[{\citenamefont{Dasgupta et~al.}(2019)\citenamefont{Dasgupta, Kang, and Popov}}]{Dasgupta:2019rmf}
\bibinfo{author}{\bibfnamefont{A.}~\bibnamefont{Dasgupta}}, \bibinfo{author}{\bibfnamefont{S.~K.} \bibnamefont{Kang}}, \bibnamefont{and} \bibinfo{author}{\bibfnamefont{O.}~\bibnamefont{Popov}}, \bibinfo{journal}{Phys. Rev. D} \textbf{\bibinfo{volume}{100}}, \bibinfo{pages}{075030} (\bibinfo{year}{2019}), \eprint{1903.12558}.

\bibitem[{\citenamefont{Enomoto et~al.}(2019)\citenamefont{Enomoto, Kanemura, Sakurai, and Sugiyama}}]{Enomoto:2019mzl}
\bibinfo{author}{\bibfnamefont{K.}~\bibnamefont{Enomoto}}, \bibinfo{author}{\bibfnamefont{S.}~\bibnamefont{Kanemura}}, \bibinfo{author}{\bibfnamefont{K.}~\bibnamefont{Sakurai}}, \bibnamefont{and} \bibinfo{author}{\bibfnamefont{H.}~\bibnamefont{Sugiyama}}, \bibinfo{journal}{Phys. Rev. D} \textbf{\bibinfo{volume}{100}}, \bibinfo{pages}{015044} (\bibinfo{year}{2019}), \eprint{1904.07039}.

\bibitem[{\citenamefont{Jana et~al.}(2019)\citenamefont{Jana, Vishnu, and Saad}}]{Jana:2019mez}
\bibinfo{author}{\bibfnamefont{S.}~\bibnamefont{Jana}}, \bibinfo{author}{\bibfnamefont{P.~K.} \bibnamefont{Vishnu}}, \bibnamefont{and} \bibinfo{author}{\bibfnamefont{S.}~\bibnamefont{Saad}}, \bibinfo{journal}{Eur. Phys. J. C} \textbf{\bibinfo{volume}{79}}, \bibinfo{pages}{916} (\bibinfo{year}{2019}), \eprint{1904.07407}.

\bibitem[{\citenamefont{Ma}(2019{\natexlab{b}})}]{Ma:2019iwj}
\bibinfo{author}{\bibfnamefont{E.}~\bibnamefont{Ma}}, \bibinfo{journal}{Eur. Phys. J. C} \textbf{\bibinfo{volume}{79}}, \bibinfo{pages}{903} (\bibinfo{year}{2019}{\natexlab{b}}), \eprint{1905.01535}.

\bibitem[{\citenamefont{Ma}(2019{\natexlab{c}})}]{Ma:2019byo}
\bibinfo{author}{\bibfnamefont{E.}~\bibnamefont{Ma}}, \bibinfo{journal}{Nucl. Phys. B} \textbf{\bibinfo{volume}{946}}, \bibinfo{pages}{114725} (\bibinfo{year}{2019}{\natexlab{c}}), \eprint{1907.04665}.

\bibitem[{\citenamefont{Restrepo et~al.}(2019)\citenamefont{Restrepo, Rivera, and Tangarife}}]{Restrepo:2019soi}
\bibinfo{author}{\bibfnamefont{D.}~\bibnamefont{Restrepo}}, \bibinfo{author}{\bibfnamefont{A.}~\bibnamefont{Rivera}}, \bibnamefont{and} \bibinfo{author}{\bibfnamefont{W.}~\bibnamefont{Tangarife}}, \bibinfo{journal}{Phys. Rev. D} \textbf{\bibinfo{volume}{100}}, \bibinfo{pages}{035029} (\bibinfo{year}{2019}), \eprint{1906.09685}.

\bibitem[{\citenamefont{Centelles~Chuli{\'a} et~al.}(2019)\citenamefont{Centelles~Chuli{\'a}, Cepedello, Peinado, and Srivastava}}]{CentellesChulia:2019xky}
\bibinfo{author}{\bibfnamefont{S.}~\bibnamefont{Centelles~Chuli{\'a}}}, \bibinfo{author}{\bibfnamefont{R.}~\bibnamefont{Cepedello}}, \bibinfo{author}{\bibfnamefont{E.}~\bibnamefont{Peinado}}, \bibnamefont{and} \bibinfo{author}{\bibfnamefont{R.}~\bibnamefont{Srivastava}}, \bibinfo{journal}{JHEP} \textbf{\bibinfo{volume}{10}}, \bibinfo{pages}{093} (\bibinfo{year}{2019}), \eprint{1907.08630}.

\bibitem[{\citenamefont{Calle et~al.}(2020)\citenamefont{Calle, Restrepo, and Zapata}}]{Calle:2019mxn}
\bibinfo{author}{\bibfnamefont{J.}~\bibnamefont{Calle}}, \bibinfo{author}{\bibfnamefont{D.}~\bibnamefont{Restrepo}}, \bibnamefont{and} \bibinfo{author}{\bibfnamefont{{\'O}.}~\bibnamefont{Zapata}}, \bibinfo{journal}{Phys. Rev. D} \textbf{\bibinfo{volume}{101}}, \bibinfo{pages}{035004} (\bibinfo{year}{2020}), \eprint{1909.09574}.

\bibitem[{\citenamefont{Kumar et~al.}(2025)\citenamefont{Kumar, Nath, Srivastava, and Yadav}}]{Kumar:2025cte}
\bibinfo{author}{\bibfnamefont{R.}~\bibnamefont{Kumar}}, \bibinfo{author}{\bibfnamefont{N.}~\bibnamefont{Nath}}, \bibinfo{author}{\bibfnamefont{R.}~\bibnamefont{Srivastava}}, \bibnamefont{and} \bibinfo{author}{\bibfnamefont{S.}~\bibnamefont{Yadav}} (\bibinfo{year}{2025}), \eprint{2505.01407}.

\bibitem[{\citenamefont{Borboruah et~al.}(2025)\citenamefont{Borboruah, Borah, Malhotra, and Patel}}]{Borboruah:2024lli}
\bibinfo{author}{\bibfnamefont{Z.~A.} \bibnamefont{Borboruah}}, \bibinfo{author}{\bibfnamefont{D.}~\bibnamefont{Borah}}, \bibinfo{author}{\bibfnamefont{L.}~\bibnamefont{Malhotra}}, \bibnamefont{and} \bibinfo{author}{\bibfnamefont{U.}~\bibnamefont{Patel}}, \bibinfo{journal}{Phys. Rev. D} \textbf{\bibinfo{volume}{112}}, \bibinfo{pages}{015022} (\bibinfo{year}{2025}), \eprint{2412.12267}.

\bibitem[{\citenamefont{De et~al.}(2022)\citenamefont{De, Das, Mitra, and Sahoo}}]{De:2021crr}
\bibinfo{author}{\bibfnamefont{B.}~\bibnamefont{De}}, \bibinfo{author}{\bibfnamefont{D.}~\bibnamefont{Das}}, \bibinfo{author}{\bibfnamefont{M.}~\bibnamefont{Mitra}}, \bibnamefont{and} \bibinfo{author}{\bibfnamefont{N.}~\bibnamefont{Sahoo}}, \bibinfo{journal}{JHEP} \textbf{\bibinfo{volume}{08}}, \bibinfo{pages}{202} (\bibinfo{year}{2022}), \eprint{2106.00979}.

\bibitem[{\citenamefont{Luo et~al.}(2020)\citenamefont{Luo, Rodejohann, and Xu}}]{Luo:2020sho}
\bibinfo{author}{\bibfnamefont{X.}~\bibnamefont{Luo}}, \bibinfo{author}{\bibfnamefont{W.}~\bibnamefont{Rodejohann}}, \bibnamefont{and} \bibinfo{author}{\bibfnamefont{X.-J.} \bibnamefont{Xu}}, \bibinfo{journal}{JCAP} \textbf{\bibinfo{volume}{06}}, \bibinfo{pages}{058} (\bibinfo{year}{2020}), \eprint{2005.01629}.

\bibitem[{\citenamefont{Luo et~al.}(2021)\citenamefont{Luo, Rodejohann, and Xu}}]{Luo:2020fdt}
\bibinfo{author}{\bibfnamefont{X.}~\bibnamefont{Luo}}, \bibinfo{author}{\bibfnamefont{W.}~\bibnamefont{Rodejohann}}, \bibnamefont{and} \bibinfo{author}{\bibfnamefont{X.-J.} \bibnamefont{Xu}}, \bibinfo{journal}{JCAP} \textbf{\bibinfo{volume}{03}}, \bibinfo{pages}{082} (\bibinfo{year}{2021}), \eprint{2011.13059}.

\bibitem[{\citenamefont{Centelles~Chuli{\'a} et~al.}(2025)\citenamefont{Centelles~Chuli{\'a}, Srivastava, and Yadav}}]{CentellesChulia:2024iom}
\bibinfo{author}{\bibfnamefont{S.}~\bibnamefont{Centelles~Chuli{\'a}}}, \bibinfo{author}{\bibfnamefont{R.}~\bibnamefont{Srivastava}}, \bibnamefont{and} \bibinfo{author}{\bibfnamefont{S.}~\bibnamefont{Yadav}}, \bibinfo{journal}{JHEP} \textbf{\bibinfo{volume}{04}}, \bibinfo{pages}{038} (\bibinfo{year}{2025}), \eprint{2409.18513}.

\bibitem[{\citenamefont{Arcadi et~al.}(2018)\citenamefont{Arcadi, Dutra, Ghosh, Lindner, Mambrini, Pierre, Profumo, and Queiroz}}]{Arcadi:2017kky}
\bibinfo{author}{\bibfnamefont{G.}~\bibnamefont{Arcadi}}, \bibinfo{author}{\bibfnamefont{M.}~\bibnamefont{Dutra}}, \bibinfo{author}{\bibfnamefont{P.}~\bibnamefont{Ghosh}}, \bibinfo{author}{\bibfnamefont{M.}~\bibnamefont{Lindner}}, \bibinfo{author}{\bibfnamefont{Y.}~\bibnamefont{Mambrini}}, \bibinfo{author}{\bibfnamefont{M.}~\bibnamefont{Pierre}}, \bibinfo{author}{\bibfnamefont{S.}~\bibnamefont{Profumo}}, \bibnamefont{and} \bibinfo{author}{\bibfnamefont{F.~S.} \bibnamefont{Queiroz}}, \bibinfo{journal}{Eur. Phys. J. C} \textbf{\bibinfo{volume}{78}}, \bibinfo{pages}{203} (\bibinfo{year}{2018}), \eprint{1703.07364}.

\bibitem[{\citenamefont{Bo et~al.}(2025)}]{PandaX:2024qfu}
\bibinfo{author}{\bibfnamefont{Z.}~\bibnamefont{Bo}} \bibnamefont{et~al.} (\bibinfo{collaboration}{PandaX}), \bibinfo{journal}{Phys. Rev. Lett.} \textbf{\bibinfo{volume}{134}}, \bibinfo{pages}{011805} (\bibinfo{year}{2025}), \eprint{2408.00664}.

\bibitem[{\citenamefont{Aalbers et~al.}(2025)}]{LZ:2024zvo}
\bibinfo{author}{\bibfnamefont{J.}~\bibnamefont{Aalbers}} \bibnamefont{et~al.} (\bibinfo{collaboration}{LZ}), \bibinfo{journal}{Phys. Rev. Lett.} \textbf{\bibinfo{volume}{135}}, \bibinfo{pages}{011802} (\bibinfo{year}{2025}), \eprint{2410.17036}.

\bibitem[{\citenamefont{Hall et~al.}(2010)\citenamefont{Hall, Jedamzik, March-Russell, and West}}]{Hall:2009bx}
\bibinfo{author}{\bibfnamefont{L.~J.} \bibnamefont{Hall}}, \bibinfo{author}{\bibfnamefont{K.}~\bibnamefont{Jedamzik}}, \bibinfo{author}{\bibfnamefont{J.}~\bibnamefont{March-Russell}}, \bibnamefont{and} \bibinfo{author}{\bibfnamefont{S.~M.} \bibnamefont{West}}, \bibinfo{journal}{JHEP} \textbf{\bibinfo{volume}{03}}, \bibinfo{pages}{080} (\bibinfo{year}{2010}), \eprint{0911.1120}.

\bibitem[{\citenamefont{Liu et~al.}(2022)\citenamefont{Liu, Guo, Zhu, and Li}}]{Liu:2022jdq}
\bibinfo{author}{\bibfnamefont{X.}~\bibnamefont{Liu}}, \bibinfo{author}{\bibfnamefont{S.-Y.} \bibnamefont{Guo}}, \bibinfo{author}{\bibfnamefont{B.}~\bibnamefont{Zhu}}, \bibnamefont{and} \bibinfo{author}{\bibfnamefont{Y.}~\bibnamefont{Li}}, \bibinfo{journal}{Sci. Bull.} \textbf{\bibinfo{volume}{67}}, \bibinfo{pages}{1437} (\bibinfo{year}{2022}), \eprint{2204.04834}.

\bibitem[{\citenamefont{Bian and Liu}(2019)}]{Bian:2018bxr}
\bibinfo{author}{\bibfnamefont{L.}~\bibnamefont{Bian}} \bibnamefont{and} \bibinfo{author}{\bibfnamefont{X.}~\bibnamefont{Liu}}, \bibinfo{journal}{Phys. Rev. D} \textbf{\bibinfo{volume}{99}}, \bibinfo{pages}{055003} (\bibinfo{year}{2019}), \eprint{1811.03279}.

\bibitem[{\citenamefont{Hessler et~al.}(2017)\citenamefont{Hessler, Ibarra, Molinaro, and Vogl}}]{Hessler:2016kwm}
\bibinfo{author}{\bibfnamefont{A.~G.} \bibnamefont{Hessler}}, \bibinfo{author}{\bibfnamefont{A.}~\bibnamefont{Ibarra}}, \bibinfo{author}{\bibfnamefont{E.}~\bibnamefont{Molinaro}}, \bibnamefont{and} \bibinfo{author}{\bibfnamefont{S.}~\bibnamefont{Vogl}}, \bibinfo{journal}{JHEP} \textbf{\bibinfo{volume}{01}}, \bibinfo{pages}{100} (\bibinfo{year}{2017}), \eprint{1611.09540}.

\bibitem[{\citenamefont{Molinaro et~al.}(2014)\citenamefont{Molinaro, Yaguna, and Zapata}}]{Molinaro:2014lfa}
\bibinfo{author}{\bibfnamefont{E.}~\bibnamefont{Molinaro}}, \bibinfo{author}{\bibfnamefont{C.~E.} \bibnamefont{Yaguna}}, \bibnamefont{and} \bibinfo{author}{\bibfnamefont{O.}~\bibnamefont{Zapata}}, \bibinfo{journal}{JCAP} \textbf{\bibinfo{volume}{07}}, \bibinfo{pages}{015} (\bibinfo{year}{2014}), \eprint{1405.1259}.

\bibitem[{\citenamefont{Guo and Han}(2020)}]{Guo:2020qin}
\bibinfo{author}{\bibfnamefont{S.-Y.} \bibnamefont{Guo}} \bibnamefont{and} \bibinfo{author}{\bibfnamefont{Z.-L.} \bibnamefont{Han}}, \bibinfo{journal}{JHEP} \textbf{\bibinfo{volume}{12}}, \bibinfo{pages}{062} (\bibinfo{year}{2020}), \eprint{2005.08287}.

\bibitem[{\citenamefont{Afanaciev et~al.}(2025)}]{MEGII:2025gzr}
\bibinfo{author}{\bibfnamefont{K.}~\bibnamefont{Afanaciev}} \bibnamefont{et~al.} (\bibinfo{collaboration}{MEG II}) (\bibinfo{year}{2025}), \eprint{2504.15711}.

\bibitem[{\citenamefont{Alloul et~al.}(2014)\citenamefont{Alloul, Christensen, Degrande, Duhr, and Fuks}}]{Alloul:2013bka}
\bibinfo{author}{\bibfnamefont{A.}~\bibnamefont{Alloul}}, \bibinfo{author}{\bibfnamefont{N.~D.} \bibnamefont{Christensen}}, \bibinfo{author}{\bibfnamefont{C.}~\bibnamefont{Degrande}}, \bibinfo{author}{\bibfnamefont{C.}~\bibnamefont{Duhr}}, \bibnamefont{and} \bibinfo{author}{\bibfnamefont{B.}~\bibnamefont{Fuks}}, \bibinfo{journal}{Comput. Phys. Commun.} \textbf{\bibinfo{volume}{185}}, \bibinfo{pages}{2250} (\bibinfo{year}{2014}), \eprint{1310.1921}.

\bibitem[{\citenamefont{Alguero et~al.}(2024)\citenamefont{Alguero, Belanger, Boudjema, Chakraborti, Goudelis, Kraml, Mjallal, and Pukhov}}]{Alguero:2023zol}
\bibinfo{author}{\bibfnamefont{G.}~\bibnamefont{Alguero}}, \bibinfo{author}{\bibfnamefont{G.}~\bibnamefont{Belanger}}, \bibinfo{author}{\bibfnamefont{F.}~\bibnamefont{Boudjema}}, \bibinfo{author}{\bibfnamefont{S.}~\bibnamefont{Chakraborti}}, \bibinfo{author}{\bibfnamefont{A.}~\bibnamefont{Goudelis}}, \bibinfo{author}{\bibfnamefont{S.}~\bibnamefont{Kraml}}, \bibinfo{author}{\bibfnamefont{A.}~\bibnamefont{Mjallal}}, \bibnamefont{and} \bibinfo{author}{\bibfnamefont{A.}~\bibnamefont{Pukhov}}, \bibinfo{journal}{Comput. Phys. Commun.} \textbf{\bibinfo{volume}{299}}, \bibinfo{pages}{109133} (\bibinfo{year}{2024}), \eprint{2312.14894}.

\bibitem[{\citenamefont{Feng et~al.}(2003)\citenamefont{Feng, Rajaraman, and Takayama}}]{Feng:2003xh}
\bibinfo{author}{\bibfnamefont{J.~L.} \bibnamefont{Feng}}, \bibinfo{author}{\bibfnamefont{A.}~\bibnamefont{Rajaraman}}, \bibnamefont{and} \bibinfo{author}{\bibfnamefont{F.}~\bibnamefont{Takayama}}, \bibinfo{journal}{Phys. Rev. Lett.} \textbf{\bibinfo{volume}{91}}, \bibinfo{pages}{011302} (\bibinfo{year}{2003}), \eprint{hep-ph/0302215}.

\bibitem[{\citenamefont{Barbieri et~al.}(2006)\citenamefont{Barbieri, Hall, and Rychkov}}]{Barbieri:2006dq}
\bibinfo{author}{\bibfnamefont{R.}~\bibnamefont{Barbieri}}, \bibinfo{author}{\bibfnamefont{L.~J.} \bibnamefont{Hall}}, \bibnamefont{and} \bibinfo{author}{\bibfnamefont{V.~S.} \bibnamefont{Rychkov}}, \bibinfo{journal}{Phys. Rev. D} \textbf{\bibinfo{volume}{74}}, \bibinfo{pages}{015007} (\bibinfo{year}{2006}), \eprint{hep-ph/0603188}.

\bibitem[{\citenamefont{Lopez~Honorez et~al.}(2007)\citenamefont{Lopez~Honorez, Nezri, Oliver, and Tytgat}}]{LopezHonorez:2006gr}
\bibinfo{author}{\bibfnamefont{L.}~\bibnamefont{Lopez~Honorez}}, \bibinfo{author}{\bibfnamefont{E.}~\bibnamefont{Nezri}}, \bibinfo{author}{\bibfnamefont{J.~F.} \bibnamefont{Oliver}}, \bibnamefont{and} \bibinfo{author}{\bibfnamefont{M.~H.~G.} \bibnamefont{Tytgat}}, \bibinfo{journal}{JCAP} \textbf{\bibinfo{volume}{02}}, \bibinfo{pages}{028} (\bibinfo{year}{2007}), \eprint{hep-ph/0612275}.

\bibitem[{\citenamefont{Cirelli et~al.}(2006)\citenamefont{Cirelli, Fornengo, and Strumia}}]{Cirelli:2005uq}
\bibinfo{author}{\bibfnamefont{M.}~\bibnamefont{Cirelli}}, \bibinfo{author}{\bibfnamefont{N.}~\bibnamefont{Fornengo}}, \bibnamefont{and} \bibinfo{author}{\bibfnamefont{A.}~\bibnamefont{Strumia}}, \bibinfo{journal}{Nucl. Phys. B} \textbf{\bibinfo{volume}{753}}, \bibinfo{pages}{178} (\bibinfo{year}{2006}), \eprint{hep-ph/0512090}.

\bibitem[{\citenamefont{Hambye et~al.}(2009)\citenamefont{Hambye, Ling, Lopez~Honorez, and Rocher}}]{Hambye:2009pw}
\bibinfo{author}{\bibfnamefont{T.}~\bibnamefont{Hambye}}, \bibinfo{author}{\bibfnamefont{F.~S.} \bibnamefont{Ling}}, \bibinfo{author}{\bibfnamefont{L.}~\bibnamefont{Lopez~Honorez}}, \bibnamefont{and} \bibinfo{author}{\bibfnamefont{J.}~\bibnamefont{Rocher}}, \bibinfo{journal}{JHEP} \textbf{\bibinfo{volume}{07}}, \bibinfo{pages}{090} (\bibinfo{year}{2009}), \bibinfo{note}{[Erratum: JHEP 05, 066 (2010)]}, \eprint{0903.4010}.

\bibitem[{\citenamefont{Aad et~al.}(2025)}]{ATLAS:2025fdm}
\bibinfo{author}{\bibfnamefont{G.}~\bibnamefont{Aad}} \bibnamefont{et~al.} (\bibinfo{collaboration}{ATLAS}), \bibinfo{journal}{JHEP} \textbf{\bibinfo{volume}{07}}, \bibinfo{pages}{140} (\bibinfo{year}{2025}), \eprint{2502.06694}.

\bibitem[{\citenamefont{Curtin et~al.}(2019)}]{Curtin:2018mvb}
\bibinfo{author}{\bibfnamefont{D.}~\bibnamefont{Curtin}} \bibnamefont{et~al.}, \bibinfo{journal}{Rept. Prog. Phys.} \textbf{\bibinfo{volume}{82}}, \bibinfo{pages}{116201} (\bibinfo{year}{2019}), \eprint{1806.07396}.

\bibitem[{\citenamefont{Mangano et~al.}(2005)\citenamefont{Mangano, Miele, Pastor, Pinto, Pisanti, and Serpico}}]{Mangano:2005cc}
\bibinfo{author}{\bibfnamefont{G.}~\bibnamefont{Mangano}}, \bibinfo{author}{\bibfnamefont{G.}~\bibnamefont{Miele}}, \bibinfo{author}{\bibfnamefont{S.}~\bibnamefont{Pastor}}, \bibinfo{author}{\bibfnamefont{T.}~\bibnamefont{Pinto}}, \bibinfo{author}{\bibfnamefont{O.}~\bibnamefont{Pisanti}}, \bibnamefont{and} \bibinfo{author}{\bibfnamefont{P.~D.} \bibnamefont{Serpico}}, \bibinfo{journal}{Nucl. Phys. B} \textbf{\bibinfo{volume}{729}}, \bibinfo{pages}{221} (\bibinfo{year}{2005}), \eprint{hep-ph/0506164}.

\bibitem[{\citenamefont{Grohs et~al.}(2016)\citenamefont{Grohs, Fuller, Kishimoto, Paris, and Vlasenko}}]{Grohs:2015tfy}
\bibinfo{author}{\bibfnamefont{E.}~\bibnamefont{Grohs}}, \bibinfo{author}{\bibfnamefont{G.~M.} \bibnamefont{Fuller}}, \bibinfo{author}{\bibfnamefont{C.~T.} \bibnamefont{Kishimoto}}, \bibinfo{author}{\bibfnamefont{M.~W.} \bibnamefont{Paris}}, \bibnamefont{and} \bibinfo{author}{\bibfnamefont{A.}~\bibnamefont{Vlasenko}}, \bibinfo{journal}{Phys. Rev. D} \textbf{\bibinfo{volume}{93}}, \bibinfo{pages}{083522} (\bibinfo{year}{2016}), \eprint{1512.02205}.

\bibitem[{\citenamefont{de~Salas and Pastor}(2016)}]{deSalas:2016ztq}
\bibinfo{author}{\bibfnamefont{P.~F.} \bibnamefont{de~Salas}} \bibnamefont{and} \bibinfo{author}{\bibfnamefont{S.}~\bibnamefont{Pastor}}, \bibinfo{journal}{JCAP} \textbf{\bibinfo{volume}{07}}, \bibinfo{pages}{051} (\bibinfo{year}{2016}), \eprint{1606.06986}.

\bibitem[{\citenamefont{Aghanim et~al.}(2020)}]{Planck:2018vyg}
\bibinfo{author}{\bibfnamefont{N.}~\bibnamefont{Aghanim}} \bibnamefont{et~al.} (\bibinfo{collaboration}{Planck}), \bibinfo{journal}{Astron. Astrophys.} \textbf{\bibinfo{volume}{641}}, \bibinfo{pages}{A6} (\bibinfo{year}{2020}), \bibinfo{note}{[Erratum: Astron.Astrophys. 652, C4 (2021)]}, \eprint{1807.06209}.

\bibitem[{\citenamefont{Abazajian et~al.}(2016)}]{CMB-S4:2016ple}
\bibinfo{author}{\bibfnamefont{K.~N.} \bibnamefont{Abazajian}} \bibnamefont{et~al.} (\bibinfo{collaboration}{CMB-S4}) (\bibinfo{year}{2016}), \eprint{1610.02743}.

\bibitem[{\citenamefont{Poulin et~al.}(2017)\citenamefont{Poulin, Lesgourgues, and Serpico}}]{Poulin:2016anj}
\bibinfo{author}{\bibfnamefont{V.}~\bibnamefont{Poulin}}, \bibinfo{author}{\bibfnamefont{J.}~\bibnamefont{Lesgourgues}}, \bibnamefont{and} \bibinfo{author}{\bibfnamefont{P.~D.} \bibnamefont{Serpico}}, \bibinfo{journal}{JCAP} \textbf{\bibinfo{volume}{03}}, \bibinfo{pages}{043} (\bibinfo{year}{2017}), \eprint{1610.10051}.

\bibitem[{\citenamefont{Hambye et~al.}(2022)\citenamefont{Hambye, Hufnagel, and Lucca}}]{Hambye:2021moy}
\bibinfo{author}{\bibfnamefont{T.}~\bibnamefont{Hambye}}, \bibinfo{author}{\bibfnamefont{M.}~\bibnamefont{Hufnagel}}, \bibnamefont{and} \bibinfo{author}{\bibfnamefont{M.}~\bibnamefont{Lucca}}, \bibinfo{journal}{JCAP} \textbf{\bibinfo{volume}{05}}, \bibinfo{pages}{033} (\bibinfo{year}{2022}), \eprint{2112.09137}.

\bibitem[{\citenamefont{Lucca et~al.}(2020)\citenamefont{Lucca, Sch{\"o}neberg, Hooper, Lesgourgues, and Chluba}}]{Lucca:2019rxf}
\bibinfo{author}{\bibfnamefont{M.}~\bibnamefont{Lucca}}, \bibinfo{author}{\bibfnamefont{N.}~\bibnamefont{Sch{\"o}neberg}}, \bibinfo{author}{\bibfnamefont{D.~C.} \bibnamefont{Hooper}}, \bibinfo{author}{\bibfnamefont{J.}~\bibnamefont{Lesgourgues}}, \bibnamefont{and} \bibinfo{author}{\bibfnamefont{J.}~\bibnamefont{Chluba}}, \bibinfo{journal}{JCAP} \textbf{\bibinfo{volume}{02}}, \bibinfo{pages}{026} (\bibinfo{year}{2020}), \eprint{1910.04619}.

\bibitem[{\citenamefont{Kawasaki et~al.}(2005)\citenamefont{Kawasaki, Kohri, and Moroi}}]{Kawasaki:2004qu}
\bibinfo{author}{\bibfnamefont{M.}~\bibnamefont{Kawasaki}}, \bibinfo{author}{\bibfnamefont{K.}~\bibnamefont{Kohri}}, \bibnamefont{and} \bibinfo{author}{\bibfnamefont{T.}~\bibnamefont{Moroi}}, \bibinfo{journal}{Phys. Rev. D} \textbf{\bibinfo{volume}{71}}, \bibinfo{pages}{083502} (\bibinfo{year}{2005}), \eprint{astro-ph/0408426}.

\bibitem[{\citenamefont{Jedamzik}(2008)}]{Jedamzik:2007qk}
\bibinfo{author}{\bibfnamefont{K.}~\bibnamefont{Jedamzik}}, \bibinfo{journal}{JCAP} \textbf{\bibinfo{volume}{03}}, \bibinfo{pages}{008} (\bibinfo{year}{2008}), \eprint{0710.5153}.

\bibitem[{\citenamefont{Escudero}(2019)}]{Escudero:2018mvt}
\bibinfo{author}{\bibfnamefont{M.}~\bibnamefont{Escudero}}, \bibinfo{journal}{JCAP} \textbf{\bibinfo{volume}{02}}, \bibinfo{pages}{007} (\bibinfo{year}{2019}), \eprint{1812.05605}.

\bibitem[{\citenamefont{Escudero~Abenza}(2020)}]{EscuderoAbenza:2020cmq}
\bibinfo{author}{\bibfnamefont{M.}~\bibnamefont{Escudero~Abenza}}, \bibinfo{journal}{JCAP} \textbf{\bibinfo{volume}{05}}, \bibinfo{pages}{048} (\bibinfo{year}{2020}), \eprint{2001.04466}.

\end{thebibliography}

\end{document}